\documentclass[times,onecolumn,final]{elsarticle}
\usepackage[svgnames]{xcolor}
\usepackage[colorlinks=true,citecolor=red]{hyperref}

\usepackage{framed,multirow}

\usepackage{amssymb}
\usepackage{latexsym}

\usepackage{url}

\usepackage{textgreek}
\usepackage{tabto}
\usepackage{tabu}
\usepackage{array, booktabs, makecell}
\usepackage{float}
\usepackage{commath}
\usepackage{amsmath}
\usepackage{amsfonts}
\usepackage{enumitem}
\usepackage{nicefrac}
\usepackage{wrapfig}
\usepackage{xcolor}
\usepackage{academicons}
\usepackage{tikz,pgfplots,adjustbox}
\usepackage{filecontents}
\usepackage{rotating}
\usepackage{pgfplotstable}

\usepackage{rotating}
\usepackage{tikz}
\usepackage{lscape}
\usepackage{sidecap}

\newcolumntype{Y}{>{\centering\arraybackslash}X}

\newcolumntype{Y}{>{\centering\arraybackslash}X}

\RequirePackage{pgf}
\RequirePackage{pgffor}
\RequirePackage{pgfpages}
\RequirePackage{tikz} 
\usetikzlibrary{
	calc,
	arrows,
	arrows.meta,
	positioning,
	matrix,
	shapes,
	shapes.geometric,
	shapes.misc,
	shapes.callouts,
	shapes.arrows,
	scopes,
	automata,
	shadows,
	fit,
	petri,
	backgrounds,
	decorations.pathreplacing,
	decorations.pathmorphing,
	decorations.markings,
	fadings,
	spy, 
	patterns
}

\usepackage{pifont}

\renewcommand{\vec}[1]{\mathbf{#1}}
\newcommand{\CNN}{\mathcal{M}}

\RequirePackage[acronym,shortcuts]{glossaries}
\glsdisablehyper

\newacronym[plural=ADs,firstplural={aortic dissections (ADs)}]{AD}{AD}{aortic dissection}
\newacronym{CTA}{CTA}{computed tomography angiography}
\newacronym{FL}{FL}{false lumen}
\newacronym{TL}{TL}{true lumen}
\newacronym{TAAD}{TAAD}{type~A aortic dissection}
\newacronym{TBAD}{TBAD}{type~B aortic dissection}
\newacronym{GAN}{GAN}{generative adversarial network}
\newacronym{LOA}{LOA}{limits of agreement}
\newacronym[plural=CNN, firstplural={convolutional neural networks (CNN)}]{CNN}{CNN}{convolutional neural network}
\newacronym{MCDS}{MCDS}{Monte Carlo dropout sampling}
\newacronym{MCDbS}{MCDbS}{Monte Carlo DropBlock sampling}
\newacronym{INS}{INS}{iterative neighbor sampling}
\newacronym{INDS}{INDS}{iterative neighbor dropout sampling}

\definecolor{newcolor}{rgb}{.8,.349,.1}

\begin{document}


\begin{frontmatter}

\title{Automated cross-sectional view selection in CT angiography of aortic dissections with uncertainty awareness and retrospective clinical annotations}

\author[1,2,3]{Antonio Pepe\corref{cor1}}
\ead{apepe@stanford.edu}
\author[3,4]{Jan {Egger}\corref{cor1}}
\ead{jan.egger@uk-essen.de}
\author[2]{Marina {Codari}}
\ead{mcodari@stanford.edu}
\author[2]{Martin J. {Willemink}}
\ead{willemink@stanford.edu}
\author[1,3]{Christina {Gsaxner}}
\ead{gsaxner@tugraz.at}
\author[3,4]{Jianning {Li}}
\ead{jianning.li@uk-essen.de}
\author[1]{Peter M. {Roth}}
\ead{pmroth@icg.tugraz.at}
\author[2,5]{Gabriel {Mistelbauer}}
\ead{gmistelbauer@isg.cs.uni-magdeburg.de}
\author[1]{Dieter {Schmalstieg}}
\ead{schmalstieg@tugraz.at}
\author[2]{Dominik {Fleischmann}}
\ead{d.fleischmann@stanford.edu}

\cortext[cor1]{Corresponding authors and shared first authorship.}

\address[1]{Graz University of Technology, Institute of Computer Graphics and Vision, Inffeldgasse 16/II, 8010 Graz, Austria.}
\address[2]{Stanford University, School of Medicine, 3D and Quantitative Imaging Lab, 300 Pasteur Drive Stanford, CA 94305, USA.}
\address[3]{Computer Algorithms for M\'edicine (Caf\'e) Laboratory, Graz, Austria.}
\address[4]{University Medicine Essen, Institute for AI in Medicine (IKIM), Girardetstraße 2, 45131 Essen, Germany.}
\address[5]{Otto-von-Guericke University. Department of Simulation and Graphics. Universitätsplatz 2, 39106 Magdeburg, Germany.}


\begin{abstract}
\emph{Objective:} 
Surveillance imaging of chronic aortic diseases, such as dissections, relies on obtaining and comparing cross-sectional diameter measurements at predefined aortic landmarks, over time. 
Due to a lack of robust tools, the orientation of the cross-sectional planes is defined manually by highly trained operators. We show how manual annotations routinely collected in a clinic can be efficiently used to ease this task, despite the presence of a non-negligible interoperator variability in the measurements. 
\emph{Impact:} Ill-posed but repetitive imaging tasks can be eased or automated by leveraging imperfect, retrospective clinical annotations.
\emph{Methodology:} 
In this work, we combine convolutional neural networks and uncertainty quantification methods to predict the orientation of such cross-sectional planes. We use clinical data randomly processed by 11 operators for training, and test on a smaller set processed by 3 independent operators to assess interoperator variability. 
\emph{Results:} 
Our analysis shows that manual selection of cross-sectional planes is characterized by 95\% limits of agreement (LOA) of $10.6^\circ$ and $21.4^\circ$ per angle. Our method showed to decrease static error by $3.57^\circ$ ($40.2$\%) and $4.11^\circ$ ($32.8$\%) against state of the art and LOA by $5.4^\circ$ ($49.0$\%) and $16.0^\circ$ ($74.6$\%) against manual processing. 
\emph{Conclusion:} 
This suggests that pre-existing annotations can be an inexpensive resource in clinics to ease ill-posed and repetitive tasks like cross-section extraction for surveillance of aortic dissections.

\end{abstract}


\end{frontmatter}


\section{Introduction}
\label{sec1}

\begin{figure*}
    \centering
    \includegraphics[width=\linewidth]{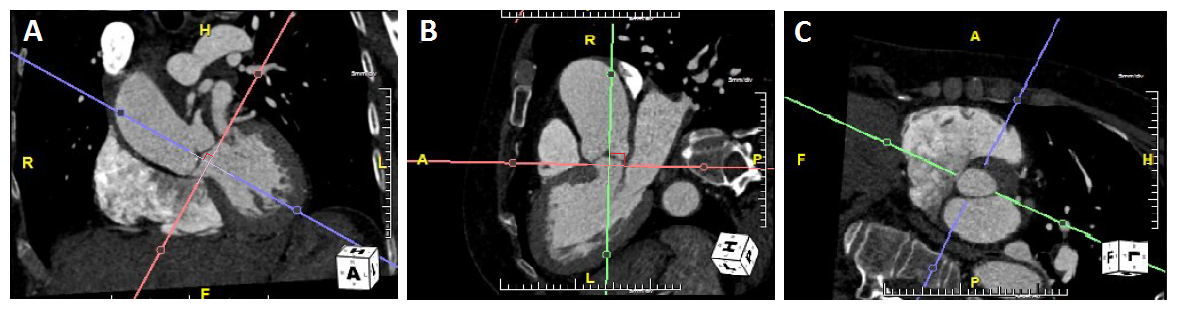}
    \caption{Example of a double-oblique reformation applied to extract the cross-sectional view of an aortic annulus.  The red planes (left and central images) are adjusted so that they are locally orthogonal to the aorta from two different orientations. The violet and green planes are parallel to the local blood flow direction. For the specific location, the orthogonal cross section is described by the red plane  (rightmost figure).}
    \label{fig:aortadoubleoblique}
\end{figure*}

\begin{figure}[tb]
    \centering
    \includegraphics[width=0.8\linewidth]{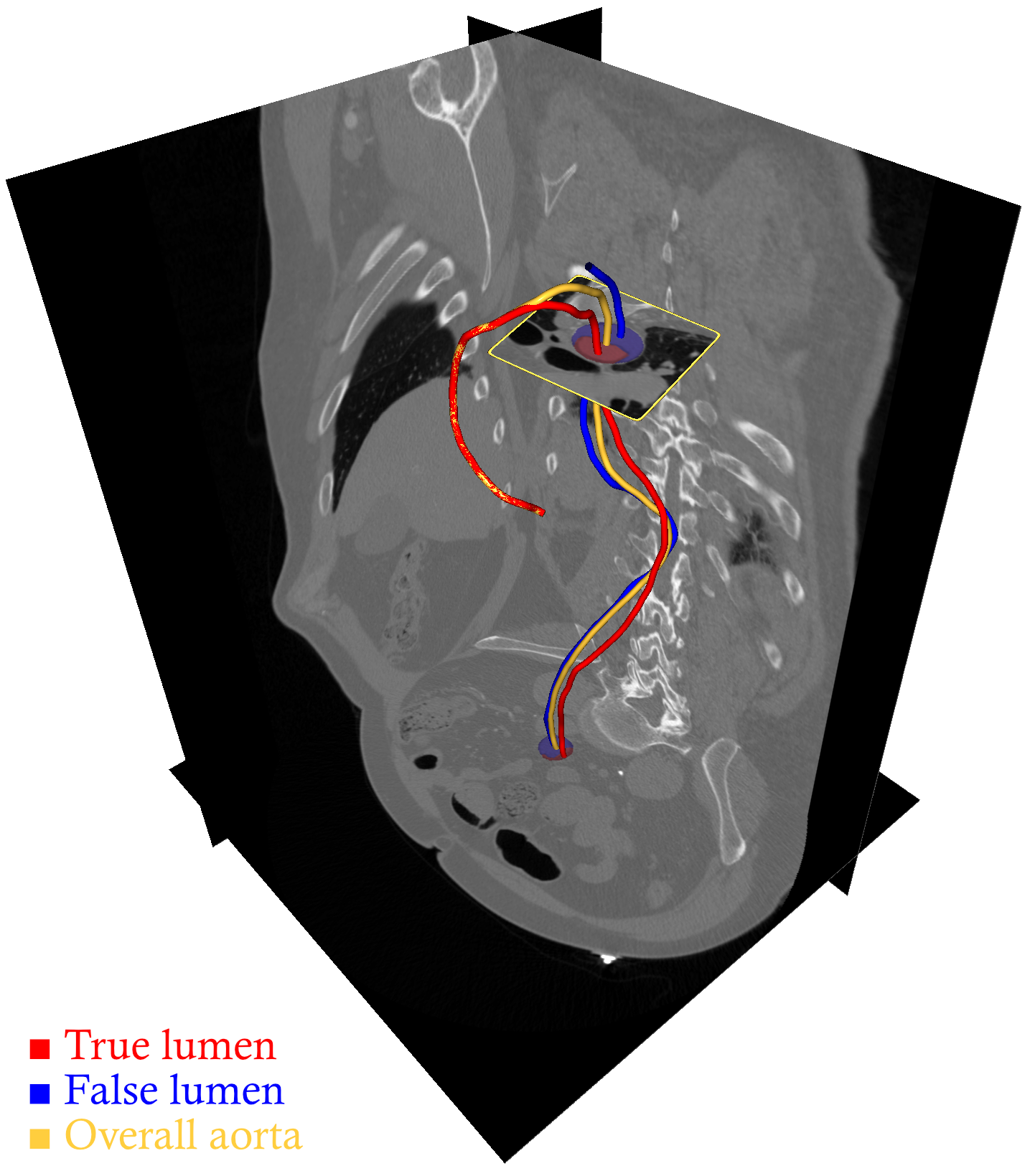}
    \caption{Examples of centerlines obtained through tinning of segmentation. The yellow line represents the centerline of the overall aorta. The direction of this centerline is used for surveillance purposes. The red and blue lines represent the true and false lumina, respectively. In this example, the false lumen originates in the descending aorta. It can be seen how the overall direction of the aorta is not necessarily matching the direction of each single lumen.  }
    \label{fig:example_centerline}
\end{figure}

\glsresetall
Cross-sectional views are regularly used for the
analysis of vascular structures in \ac{CTA} images~(\cite{mueller_eschner_2013, Rajiah_2013, guidelines_2014, Gamechi_2019}). Patients with chronic aortic diseases, such as aneurysms and dissections, require life-long surveillance and serial imaging to detect aneurysm growth and prevent fatal aortic rupture~(\cite{Lau_2017}). The most important information sought in aortic surveillance imaging is aortic caliber. An accurate determination of aortic caliber requires that diameter measurements are obtained on cross-sections oriented orthogonal to the aortic flow channel~(\cite{guidelines_2014,Lombardi_2020,Pepe_2020}).
In clinical practice, the selection and orientation of these cross-sectional views, or planes, is defined manually by highly trained radiologists and radiology technicians, using free-hand interactive double-oblique reformations~(\cite{diazpelaez_2017,bahve2018}) or semi-automatic techniques that track the vessel centerline~(\cite{mueller_eschner_2013,Gamechi_2019}). \autoref{fig:aortadoubleoblique} and \autoref{fig:example_centerline} provide examples of these methodologies.

\begin{figure*}
    \centering
    \includegraphics[width=0.9\linewidth]{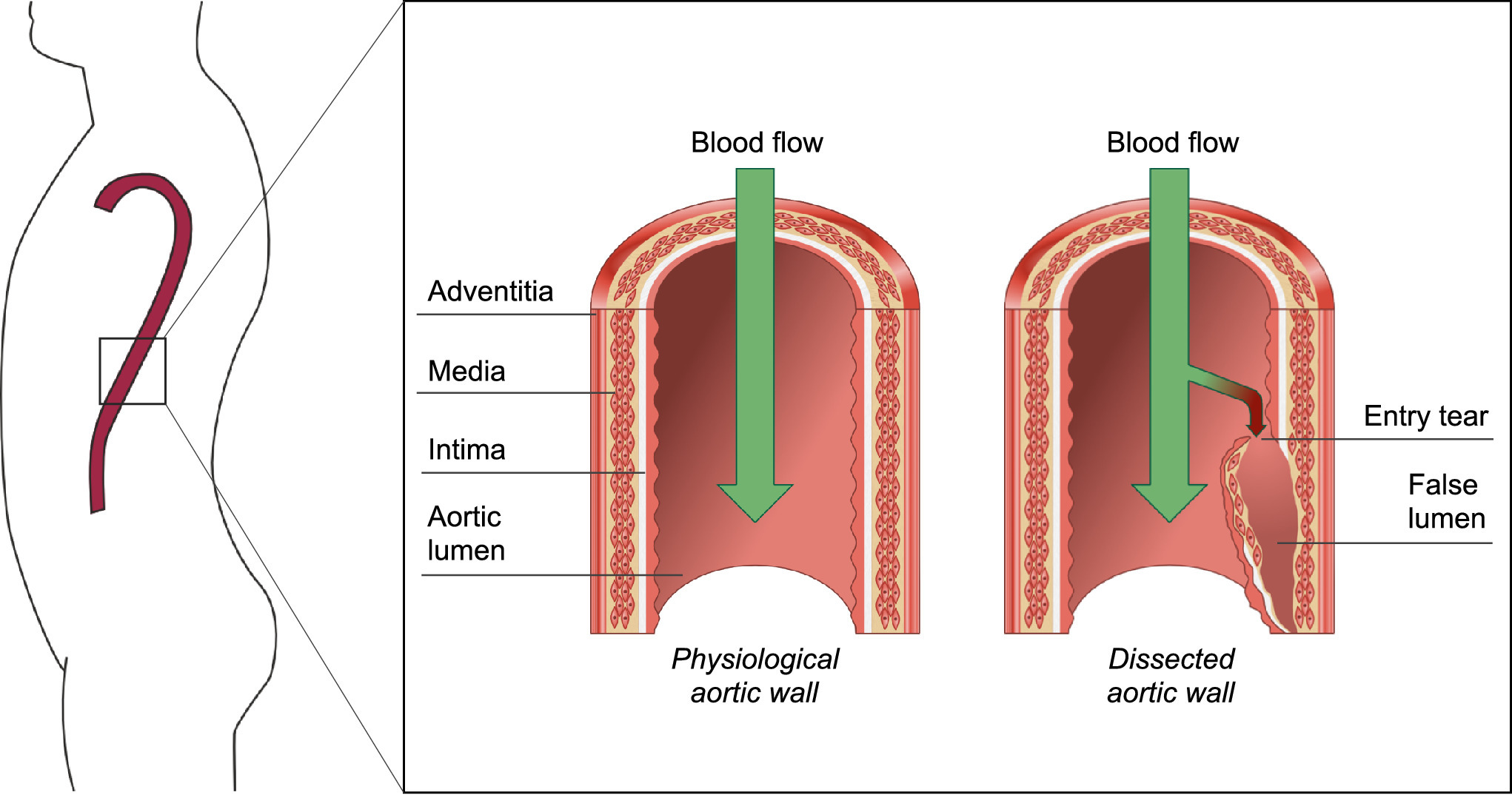}
    \caption{Computer-generated visualization of the aortic wall, of its different layers, and of its deformation due to the formation of an entry tear~(\cite{Pepe_2020}).}
    \label{fig:illustration_ad}
\end{figure*}

\begin{figure}
    \centering
    \includegraphics[width=\linewidth]{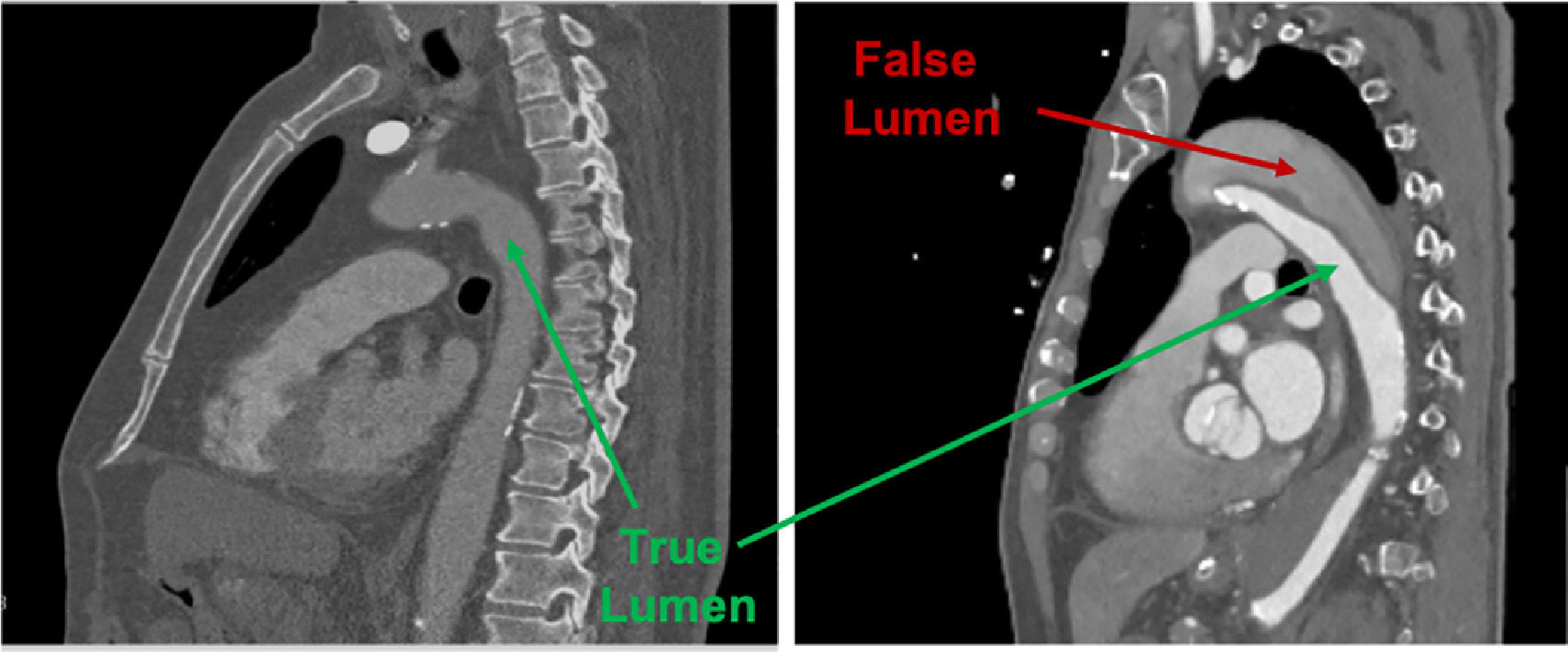}
    \caption{Computed tomography angiography of a healthy subject (left), where only one lumen is visible in the aorta; an aortic dissection case (right), where the false and true lumina are clearly distinguishable~(\cite{Pepe_2020}).}
    \label{fig:healthy_diss}
\end{figure}

The manual, free-hand estimation of these cross-sectional views and measurements is not only time-consuming, requiring about $15$ minutes per patient, but the resulting measurements are also operator-dependent: 
\cite{Nienaber_2016} showed that the interoperator and intraoperator variabilities for cross-sectional diameter measurements in CTA images of aortic aneurysms and dissections are $\pm 5$\,mm and $\pm 3$\,mm, respectively.
This introduces a considerable uncertainty in clinical decision making, where indications for surgical repair are based on a measurement threshold of $55$\,mm~(\cite{heuts_2020}).
The centerline-based approaches have been validated in patients with patent and unaltered flow channels~(\cite{Gamechi_2019}). However, just the presence of eccentric thrombus within an aneurysm introduces significant inaccuracies in centerline-based diameter measurements~(\cite{Egger_2012,Krissian_2014,Kaufhold_2018}).
The morphologic changes related to \ac{AD}, where a second blood-flow channel forms between the layers of the aortic wall, often prohibit the use of automated centerline extraction methods~(\cite{Pepe_2020}). \autoref{fig:illustration_ad} and \autoref{fig:healthy_diss} show the differences between a healthy aortic lumen and a case of \ac{AD}, where also a second flow channel (false lumen) is present. 
The flow channels are typically asymmetric, often with different contrast medium opacification (i.e. voxel intensity), tortuous and twisting around each other (\autoref{fig:healthy_diss}, right), and both -- the original true lumen and the new, so-called false lumen of the dissected aorta -- can contain thrombus. Thus, in patients with aortic dissection, the time consuming and operator-dependent manual selection is standard of care. The existence of clinical-grade labels -- while recognizing and accounting for their known limitations -- provide an opportunity to pursue an alternative approach, improving the speed and reproducibility of aortic measurements~(\cite{Houben_2020}).

\paragraph{Contribution}
In this work, we show how an uncertainty-aware \ac{CNN} can be trained on imperfect annotations obtained in a routine clinical setting, to extract cross-sections with higher reproducibility, and within the accuracy of current clinical practice standards. In particular, we:
\begin{itemize}
    \item provide an overview of the clinical needs and the potential of data-driven methods, such as deep learning~(\autoref{sec:tech_background}),
    \item formulate our research hypothesis and propose an uncertainty-aware deep learning method for the automation of the clinical work based on existing clinical annotations~(\autoref{sec:method}),
    \item compare the proposed method with other state-of-the-art algorithms and with the interoperator variability of three independent experts~(\autoref{sec:results}), and
    \item discuss the role and impact of uncertainty quantification for the utilization of raw clinical annotations from different experts~(\autoref{sec:discuss}).
\end{itemize}
To the best of our knowledge, this is the first investigation of uncertainty-aware deep learning approaches for the extraction of cross-sectional views in \ac{CTA} images with a chronic aortic disease, for which the common tubular prior does not hold.

\section{Background}
\label{sec:tech_background}
In this section, we provide a general overview of the pathology (\autoref{pathology}) and the current clinical surveillance protocols (\autoref{surveillance}). Readers with a strong cardiovascular background might want to skip this initial overview. Afterwards, we analyze current state-of-the-art algorithms for the image processing (\autoref{sota}) and discuss the impact that uncertainty-aware deep learning algorithms can have (\autoref{potential_DL}). 



\subsection{Pathology}
\label{pathology}
The aorta is the main artery in the human body. Microscopically, the aortic wall consists of three layers: a thin intima, a thick and elastic media, an outer-most and fibrous adventitia~(\cite{Sherifova_2019}). Aortic dissection is characterized by a splitting and delamination of these layers resulting in a new flow channel -- the \ac{FL} -- which is separated from the original \ac{TL} by the delaminated portion of the aortic wall, called the dissection flap~(\autoref{fig:illustration_ad})~(\cite{Sherifova_2019}). The origin of this delamination is referred to as \textit{entry tear}, which creates the communication between \ac{TL} and \ac{FL}, and \emph{may} be followed by one or more \textit{re-entry tears}, which allow blood to flow back to the \ac{TL}. \ac{AD} weakens the aortic wall and may cause a range of life-threatening complications~(\cite{slonim1996aortic}).

\begin{figure*}[t]
    \centering
    \includegraphics[width=\linewidth]{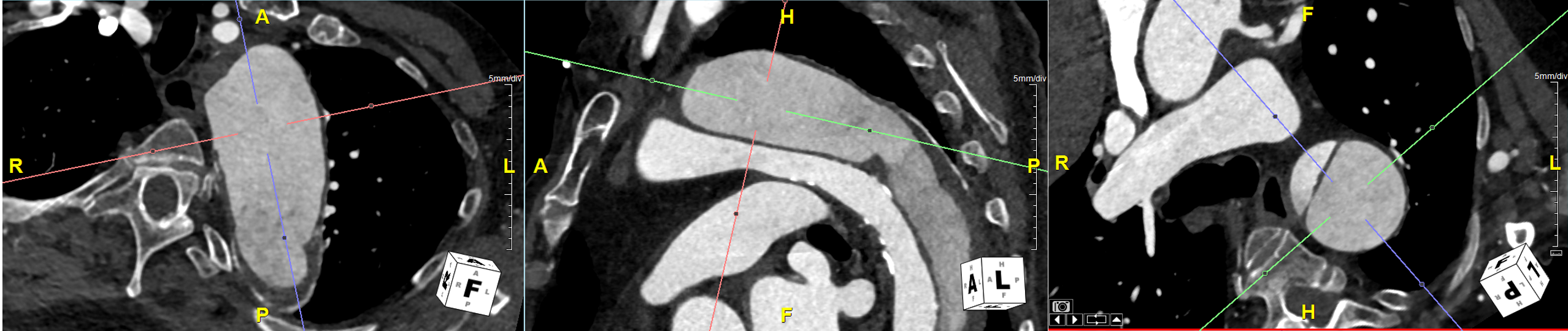}
    \caption{Example of a double-oblique reformation applied to a case of aortic dissection. In the central and right images it is possible to distinguish the two lumina of a dissected aorta. The presence of two channels introduces further uncertainty in the manual task.  The red planes (left and central images) are adjusted so that they are orthogonal to the aorta from two different orientations. The violet and green planes are parallel to the blood flow direction. For the specific location, the orthogonal cross section is described by the red plane (rightmost figure).}
    \label{fig:aortadoubleobliqueAD}
\end{figure*}

\subsection{Clinical surveillance}
\label{surveillance}
Patients with \ac{AD} undergo \ac{CTA} imaging for their initial diagnosis, and for the rest of their lives after hospital discharge, typically after 3 months, 6 months, 12 months, and annually thereafter. Surveillance imaging includes orthogonal diameter measurements along the aorta at specific landmark positions~(\cite{mueller_eschner_2013,Mistelbauer_2016,diazpelaez_2017,Nienaber_2016,bahve2018,Cao_2019,hahn_2020,Houben_2020}). A diameter larger than 55~mm is usually an indicator for eligible surgery~(\cite{heuts_2020}). This threshold was determined based on the risk of aortic rupture~(\cite{Nienaber_2016}).
Examples of additional indicators are quality of arterial perfusion and overall extension of the \ac{FL}, to exclude the imminent risk of ischemic complications~(\cite{slonim1996aortic}).

After \ac{CTA} acquisition, aortic diameters are measured in orthogonal views obtained through a reformation process~(\cite{kauffmann_2011_do, Nienaber_2016,diazpelaez_2017}). Examples of reformation techniques are the manual, double-oblique reformation~(\cite{kauffmann_2011_do}) and the curved multi-planar reformation~(\cite{Kanitsar_2002,hahn_2020}). The double-oblique reformation consists in visually determining a new coordinate system for which the XY plane appears orthogonal to the specific location of the aorta, while the planes YZ and XZ are parallel to aortic wall direction in the sagittal and coronal views, respectively (\autoref{fig:aortadoubleobliqueAD})~(\cite{kauffmann_2011_do}). To automate this process, curved multi-planar reformations were introduced~(\cite{Kanitsar_2002,hahn_2020}). In this case, the cross-sectional views are considered as orthogonal to the centerline of the aorta. It is important to distinguish the centerline of the overall aorta, used for clinical surveillance~(\cite{hahn_2020}), from the centerline of the \ac{TL}, which is used for surgical planning~(\cite{zhao2021automatic}). A comparative example is shown in \autoref{fig:example_centerline}. In this work, we focus on applications for clinical surveillance.

\subsection{\ac{AD} image analysis}
\label{sota}
Technically, the previously mentioned centerlines can be obtained by applying tracking filters~(\cite{Krissian_2014}) or through thinning of the segmentation, i.e., skeletonization~(\cite{hahn_2020}).
However, clinical validation showed that diameter measurements obtained through multi-planar reformations tend to fail or overestimate the aortic diameter in patients with aortic dissection or aneurysm, especially in presence of an aortic thrombus, where tracking methods fail~(\cite{kauffmann_2011_do,Krissian_2014}), or, track the two lumina separately~(\cite{Krissian_2014}). Only recently, \ac{CNN}-based segmentation approaches showed higher robustness also in presence of thrombus~(\cite{Cao_2019,hahn_2020,cheng2020deep,chen2021multi}).

These recent methods provided similar accuracy, although \cite{chen2021multi} were also able to simultaneously segment the major arteries. For our purposes, the evaluation of \cite{hahn_2020} is of special interest. The authors specifically evaluated their method on cross-sectional view selection. Their \ac{CNN} was trained and evaluated on 153 manually labelled \ac{CTA} volumes and can be used to segment the two lumina of a dissected aorta in presence of thrombus. They used the segmentation of the two lumina to compute the centerline of the overall aorta and retrieve the cross-sectional planes.
Although this approach delivers a valid solution, the accurate manual labelling of a conspicuous number of volumes is often not viable in small and medium-sized clinics. An experienced radiologist requires one to two hours to manually label one dissected aorta, therefore, \textit{preparing $150$ cases requires two months of full-time work, excluding review and revision}. Public collections with labelled volumes can often be a solution to support research and clinical centers, but these are difficult to obtain for each different aortic pathology, and the quality of the imaging data could differ from clinic to clinic~(\cite{Pepe_2020}). We expect this effort to be comparable to that of the other studies, although a complete description is not available.

\subsection{Uncertainty-aware deep learning}
\label{potential_DL}
Deep learning showed impressive results over the last years, also in medical image analysis~(\cite{shen2017deep}). However, its clinical application is limited by its lack of robustness. Recent reviews and editorials on deep learning are explicitly pointing out to the importance of task reproducibility in medicine~(\cite{Park2019ReprodRadiom,Stupple2020ReprodCrisis,HaibeKains2020Reprod}). It has also been pointed out how the high accuracy of a (deep learning) model does not guarantee the validity of the model for real-world applications~(\cite{Li2020AccuracyReprod}). Therefore, it is essential to define supervision strategies that can help a user monitor patient-level predictions and possibly correct them, especially when reliability and reproducibility are key factors~(\cite{begoli2019need,jungo2019assessing}). 
An important factor for model validation is the robustness against different, and likely imperfect, references and, therefore, against a non-negligible data uncertainty (e.g., quality of the input). This translates in expecting not only a high accuracy, but also a low model uncertainty. Additionally, this uncertainty can be used to support decision-making processes or to enable fast and effective corrections of the computed results. Recent works started to investigate this aspect, particularly for image segmentation tasks~(\cite{jungo2019assessing,wang2019aleatoric}).

Convolutional neural networks have found a number of applications for predictive tasks in biomedical imaging~(\cite{xie_2015}). Yet, their coupling with data and model uncertainty has been less intensively investigated. \cite{kendall_2017} evaluated the impact of data and model uncertainty in computer vision applications using Bayesian neural networks. {They} showed how model (or epistemic) uncertainty becomes less relevant compared to data uncertainty when the model is trained on large data collections. Vice versa, model uncertainty becomes critical when the model is trained on smaller collections. This is relevant in medical image analysis, where it is difficult to collect large amounts of labelled data for rare diseases. Classical neural networks do not quantify model uncertainty, and Bayesian neural networks, which describe the weight parameters as distributions over the parameters $\omega_i = f(\mu_i,\sigma_i)$~(\cite{kendall_2017}), can be harder to train, especially in 3D medical imaging, where data can be scarce and requires higher computational costs~(\cite{gal_2016,li2021automatic}). An alternative approach is to train $K$ models independently. Given the data pairs $(\vec{x}, \vec{y})$, and the weights' distribution $\mathcal{D}$, the output distribution of the model $p(\vec{y}|\vec{x},\mathcal{D})$ can be represented, empirically, as the mean and variance of the $K$ independent predictions. This can be achieved, for example, by training the $K$ models using bootstrapping~(\cite{Lakshminarayanan_2016}). Alternatively, \cite{gal_2016} proposed \ac{MCDS}. The authors apply dropout, a common regularization technique used during training~(\cite{dropout}), also during testing. The dropout regularization deactivates a random subset of neurons, implicitly generating a pool of models. Some initial applications of \ac{MCDS} for \ac{CNN} detection and segmentation have been recently reported~(\cite{leibig_2017,eaton_rosen_2018,Ayhan_2020,nair_2020}). These studies provided promising results, but some effects of dropout uncertainty quantification for a \ac{CNN} need to be further investigated: The sparse and random deactivation of a neuron brings lower benefits with convolutional layers due to the spatial relation built by the convolutional kernels~(\cite{DropBlock}). For a broader overview on deep learning and uncertainty quantification in medical imaging, we refer the reader to recent reviews and contributions from \cite{abdar2021review}, \cite{ghesu2021quantifying}, and \cite{ghoshal2021estimating}. Following these considerations, we investigate how these methods can be coupled with clinical annotations in the remainder of this work.

\begin{figure}[t]
    \centering
    \begin{tikzpicture}[
            box/.style n args={2}{
                anchor=west,
				inner sep=0pt,
				draw=black,
				fill=white,
				rectangle,
				minimum width=#1 cm,
				minimum height=#2 cm,
				font=\ttfamily\large,
            },
            putimage/.style n args={3}{
                box={#1}{#2},
                draw=black!20,
                path picture={
            	    \node[anchor=north] at (path picture bounding box.north) {
            	    	\includegraphics[height=#2 cm]{#3}
                	};
                },
            },
            putimageNoEdge/.style n args={3}{
                box={#1}{#2},
                draw=black!0,
                path picture={
            	    \node[anchor=north] at (path picture bounding box.north) {
            	    	\includegraphics[height=#2 cm]{#3}
                	};
                },
            },
            boxlink/.style={
            	black,
            	line width=1pt,
            	-{Triangle[length=5pt, width=3pt]},
            },
        ]
        
        \pgfmathsetmacro{\myW}{3.8}
        \pgfmathsetmacro{\mySep}{0.55}
       
    
        \node (I-1) [putimage={2}{\myW}{figures/w0}] {};
        \node (I-2) [right=\mySep of I-1.east,putimage={\myW}{\myW}{figures/click_3d}] {};
        \node (CNN) [right=\mySep of I-2.east,putimageNoEdge={0.65}{0.65}{figures/cnn}] {};
        \node (I-3) [right=\mySep of CNN.east,putimage={\myW}{\myW}{figures/click_3d_res}] {};
        
        
        \draw[boxlink] (I-1) -- (I-2);
        \draw[boxlink] (I-2) -- (CNN);
        \draw[boxlink] (CNN) -- (I-3);
    \end{tikzpicture}
    \caption{%
        Illustration of the processing pipeline.
        From left to right: The user selects the location of interest in a 3D volume; the system extracts a 3D patch centered at the specified location; the proposed deep learning architecture predicts the cross-sectional plane.
        Although we refer to raw \ac{CTA} data, only the aorta is shown for simplicity. In particular, we illustrate a dissected aorta, where two flow channels are present, separated by a flap. The flap limits the applicability of centerline tracking algorithms.
    }
    \label{fig:scheme}
\end{figure}
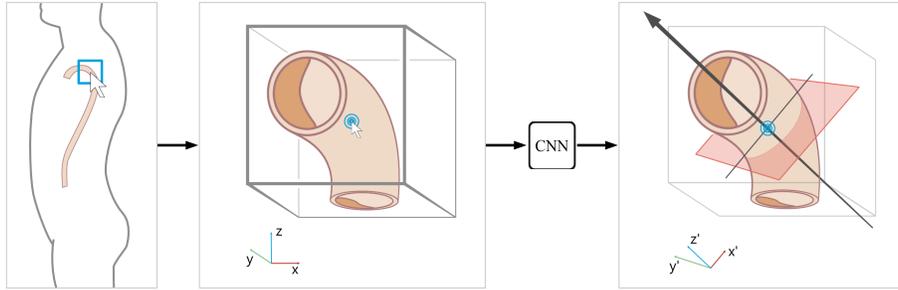

\section{Method}
\label{sec:method}
In this section, we formulate our research hypothesis and expand it into an uncertainty-aware \ac{CNN} approach for the extraction of the cross-sectional planes. 

\subsection{Research hypothesis} In the previous section, we discussed the cost of image annotation for \ac{AD}. An alternative source of data might come from clinical annotations. The routinely performed manual selections of double-oblique reformations in aortic surveillance produces an increasing amount of annotated \ac{CTA} volumes without additional costs, given the high diffusion of cardiovascular diseases and the need for follow-up examinations at regular time intervals~(\cite{Mistelbauer_2016}). 
This sets the ground for the definition and evaluation of specific data-driven approaches, which can leverage the availability of such an amount of data. This can result in less expensive solutions, which need, however, to take into account some important factors: The aortic locations at which the measurements are taken may vary from clinic to clinic and from patient to patient; both the plane selection and the measurements are operator-dependent; some less common diseases might be under-represented and more strongly affected by interoperator variations. Moreover, depending on the disease, the sampled data may be scarce or incomplete, and it may show a non-negligible interoperator variability, suggesting the application of uncertainty-aware machine learning models~(\cite{begoli2019need}).

\begin{figure}
    \centering
    \begin{tikzpicture}[
            box/.style n args={2}{
                anchor=west,
			    inner sep=0pt,
				draw=black,
				fill=white,
				rectangle,
				minimum width=#1 cm,
				minimum height=#2 cm,
				font=\ttfamily\large,
            },
            putimage/.style n args={3}{
                box={#1}{#2},
                draw=black!20,
                path picture={
            	    \node[anchor=north] at (path picture bounding box.north) {
            	    	\includegraphics[height=#2 cm]{#3}
                	};
                },
            },
            putimageNoEdge/.style n args={3}{
                box={#1}{#2},
                draw=black!0,
                path picture={
            	    \node[anchor=north] at (path picture bounding box.north) {
            	    	\includegraphics[height=#2 cm]{#3}
                	};
                },
            },
            boxlink/.style={
            	black,
            	line width=1pt,
            	-{Triangle[length=5pt, width=3pt]},
            },
        ]
        
        \pgfmathsetmacro{\mySep}{0.52}
       
    
        \node (I-1) [putimage={3.5}{2.1}{figures/click_2d}] {};
        \node (CNN) [right=\mySep of I-1.east,putimageNoEdge={0.85}{0.85}{figures/cnn}] {};
        \node (I-2) [right=\mySep of CNN.east,putimage={3.5}{2.1}{figures/click_2d_res}] {};
        
        
        \draw[boxlink] (I-1) -- (CNN);
        \draw[boxlink] (CNN) -- (I-2);
        
        \draw[boxlink] ($(I-2.west |- CNN.south)+(down:\mySep)$) -| (CNN.south);
    \end{tikzpicture}
    \caption{%
        Illustration of the iterative neighbor sampling technique.
        The user provides an initial pivot point (seed); an iterative process evaluates the robustness of the result by analyzing the plane prediction of neighbor voxels, which lie on the predicted plane.
    }
    \label{fig:iterative_sampling}
\end{figure}
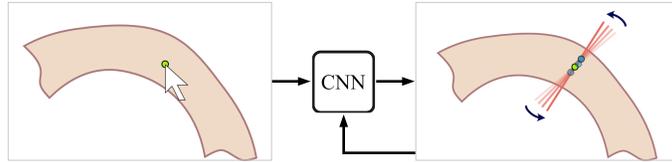

\subsection{Model architecture}
\label{sec:archit}
In recent years, convolutional neural networks were applied for regression tasks~(\cite{xie_2015}). A typical architecture for regression uses convolutional encoding layers to extract meaningful latent features and fully connected layers to map these features to the desired outputs. A limitation of such an architecture is its inability to represent model uncertainty. Recent studies added dropout layers to the convolutional layers to perform uncertainty quantification through \ac{MCDS}~(\cite{leibig_2017,nair_2020}). 
However, current applications of ~\ac{MCDS} use dropout regularization with all layers, although this was initially only conceived for fully connected layers~(\cite{nair_2020,DropBlock}). 
We illustrate our \ac{CNN} architecture in~\autoref{fig:conv_network}:
In the encoding layers (first nine blocks from the left), we use batch normalization and DropBlock~(\cite{DropBlock}) to regularize the \ac{CNN}, max pooling and convolution stride to reduce the number of relevant features. Unlike recent works~(\cite{leibig_2017,nair_2020}), we chose to replace the dropout layers with DropBlock regularization in the encoding layers. This allows us to obtain a Bayesian approximation through Monte Carlo sampling and perform an efficient regularization at the same time. The random deactivation of whole neighborhoods has been shown to provide better regularization in \ac{CNN}~(\cite{DropBlock}).
The encoding layers are followed by fully connected layers, which perform the regression from the encoded latent space to the spherical coordinates $(\theta, \phi)$ of the normal vector.
Dropout is still applied as a regularization technique on the fully connected layers.
As input, for each pivot point $P_i$, we extract its surrounding patch of size $64\times64\times64$.

\subsection{Training strategy}
To prevent overfitting, we augment the training set with random translations.
The translation is achieved by considering additional pivot points, which lie on the annotated cross-sectional plane $\Pi_i$.
These points are collected from a uniform 2D distribution centered in $P_i$ with a radius of $r = 1.53$\,mm, which is the standard deviation for $95\%$ LOA of up to $\pm3.0$\,mm. This guarantees higher robustness to inaccuracies during the placement of the pivot points (see~\autoref{fig:landmark_pos}). We rely on a 2D distribution as this guarantees that the plane orientations are still correct -- orthogonal displacements might require to remeasure the cross-sectional orientations, especially in the ascending aorta or in patients with higher tortuosity. For the training of the network, we employ a batch size of $b = 64$, a learning rate of $l_r = 0.002$, Adam optimizer~(\cite{Kingma_2015}), and early stopping. Additionally, due to the periodic nature of the data at hand, we define and apply a circular Huber loss. The standard Huber loss is a piecewise function: 
\begin{align}
    H(x,y) &= \begin{cases} 
          \frac{1}{2}( x - y )^2, & | x - y | < \delta \\
          \delta| x - y | - \frac{1}{2}\delta^2, & \text{otherwise}, \\
       \end{cases}
\end{align}
where we set $\delta = 1$, as suggested by \cite{Ross2015}.

Although Huber loss has shown to be more robust to outliers than more common losses like $\mathcal{L}_1$ or $\mathcal{L}_2$~(\cite{Ross2015}), it was not meant for periodic functions, such as angles. 
We defined a circular Huber loss to consider only the acute angles between targets and predictions: 
\begin{align}
    H_{C}(\alpha,\beta) &= \begin{cases}
        H(\alpha, \beta), & | \alpha - \beta | \leq \frac{\pi}{2} \\
        H(x, \beta - sign(\beta) \; \pi), & \text{otherwise}.
    \end{cases}
\end{align}
It takes into account that the geometrical layout for the reference angles will always be within the closed interval $[-\frac{\pi}{2},\frac{\pi}{2}]$ and that two angles $\frac{\pi}{2}+k$ and $-\frac{\pi}{2}+k$ describe the same view, i.e., our function is periodic with period $\pi$.

\begin{landscape}

\begin{figure}[t]
    \centering
    \begin{turn}{-180}
    \definecolor{myCol1}{RGB}{251,180,174}
    \definecolor{myCol2}{RGB}{179,205,227}
    \definecolor{myCol3}{RGB}{204,235,197}
    \definecolor{myCol4}{RGB}{222,203,228}
    \begin{tikzpicture}[
            box/.style n args={2}{
				fill=#2,
				rectangle,
				minimum width=#1 cm,
				minimum height=0.5cm,
				text width=#1 cm,
				anchor=north,
				rotate=90,
				align=center,
				font=\ttfamily\fontsize{8}{10}\selectfont,
			},
			boxlink/.style={
				black,
				line width=1pt,
				-{Triangle[length=5pt, width=3pt]},
			},
        ]
        
        \pgfmathsetmacro{\mySep}{0.265}
        
        \node (L-1A) [box={3.4}{black!15},draw=none] {DropBlock(3\textsuperscript{3}, $\frac{1}{5}$)};
        \node (L-1) [right=\mySep of L-1A.south,box={3.4}{black!10},draw=black,fill=none] {Conv(7\textsuperscript{3}, 32, 1) \\ Batch Norm, ReLU};
        \node (L-1B) [right=\mySep of L-1.south,box={3.4}{black!15},draw=none] {DropBlock(3\textsuperscript{3}, $\frac{1}{5}$)};
        \node (L-2) [right=\mySep of L-1B.south,box={3.4}{black!10},draw=black,fill=none] {Conv(3\textsuperscript{3}, 64, 2) \\ Batch Norm, ReLU \\ MaxPool(3, 2)};
        \node (L-2B) [right=\mySep of L-2.south,box={3.4}{black!15}] {DropBlock(3\textsuperscript{3}, $\frac{3}{20}$)};
        \node (L-3) [right=\mySep of L-2B.south,box={3.4}{black!10},draw=black,fill=none] {Conv(3\textsuperscript{3}, 128, 2) \\ Batch Norm, ReLU \\ MaxPool(3, 1)};
        \node (L-3B) [right=\mySep of L-3.south,box={3.4}{black!15}] {DropBlock(3\textsuperscript{3}, $\frac{1}{10}$)};
        \node (L-0) [right=\mySep of L-3B.south,box={3.4}{black!10},draw=black,fill=none] {Conv(3\textsuperscript{3}, 256, 2) \\ Batch Norm, ReLU};
        \node (L-0B) [right=\mySep of L-0.south,box={3.4}{black!15}] {DropBlock(3\textsuperscript{3}, $\frac{1}{10}$)};
        \node (L-4) [right=\mySep of L-0B.south,box={3.4}{black!10}] {FC(2048, 1024) \\ ReLU};
        \node (L-5) [right=\mySep of L-4.south,box={3.4}{black!15}] {Dropout($\frac{7}{20}$)};
        \node (L-6) [right=\mySep of L-5.south,box={3.4}{black!10}] {FC(1024, 1024) \\ ReLU};
                \node (L-7) [right=\mySep of L-6.south,box={3.4}{black!15}] {Dropout($\frac{1}{5}$)};
        \node (L-8) [right=\mySep of L-7.south,box={3.4}{black!10}] {FC(1024, 256) \\ ReLU};
        \node (L-9) [right=\mySep of L-8.south,box={3.4}{black!10}] {FC(256, 2)};
        
        
        \draw[boxlink] ($(L-1A.north)+(left:\mySep)$) node[left,rotate=90,xshift=1.4cm,yshift=0.3cm,text width=2.5cm,font=\ttfamily\fontsize{8}{10}\selectfont,align=center] {\textbf{image patch}} -- (L-1A);
         \draw[boxlink] (L-1A) -- (L-1);
        \draw[boxlink] (L-1) -- (L-1B);
        \draw[boxlink] (L-1B) -- (L-2);
        \draw[boxlink] (L-2) -- (L-2B);
        \draw[boxlink] (L-2B) -- (L-3);
        \draw[boxlink] (L-3) -- (L-3B);
        \draw[boxlink] (L-3B) -- (L-0);
        \draw[boxlink] (L-0) -- (L-0B);
        \draw[boxlink] (L-0B) -- (L-4);
        \draw[boxlink] (L-4) -- (L-5);
        \draw[boxlink] (L-5) -- (L-6);
        \draw[boxlink] (L-6) -- (L-7);
        \draw[boxlink] (L-7) -- (L-8);
        \draw[boxlink] (L-8) -- (L-9);
        \draw[boxlink] (L-9) -- ($(L-9.south)+(right:\mySep)$) node[left,rotate=90,xshift=0.67cm,yshift=-0.3cm,text width=1.cm,font=\ttfamily\fontsize{8}{10}\selectfont,align=center] {$\boldsymbol{(\theta, \phi)}$};
    
    \end{tikzpicture}
   \end{turn}

    \caption{Illustration of the \ac{CNN} architecture. Legend: DropBlock(K, R) -- DropBlock layer with kernel size K and drop rate R; Conv(K, D, S) -- convolutional layer with kernel size K, stride S, and D output channels; FC(X, Y) -- fully connected layer with X inputs and Y outputs; Dropout(R) -- Dropout layer with drop rate R.}

    \label{fig:conv_network}
\end{figure}
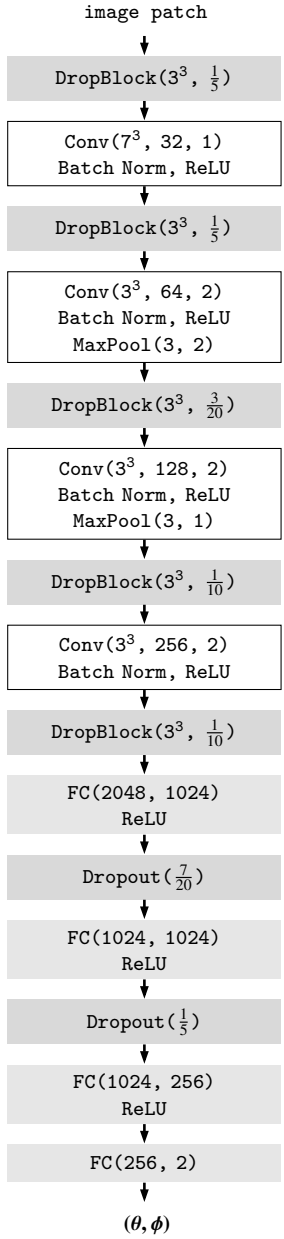

\end{landscape}

\subsection{Execution strategy}
\label{sec:exec_strategy}
Interoperator variability of expert annotations is not negligible~(\cite{Nienaber_2016}). It is therefore important to leverage this information of variability during plane selection. A classic \ac{CNN} \color{black}generates a prediction for a given input without providing any information about the uncertainty on the prediction itself. In \autoref{sec:tech_background}, we discussed the main approaches for model uncertainty quantification and some of their limitations. Here, we define and evaluate two execution strategies, which extend the concept of \ac{MCDS}~(\cite{gal_2016}). We refer to these strategies as \ac{INS} and \ac{MCDbS}.

\paragraph*{\Ac{MCDS}} Initially suggested by \cite{gal_2016}, \ac{MCDS} uses the variability produced by the dropout layers to perform a Bayesian approximation through multiple executions.
Dropout layers are generally deactivated after the test phase.
In \ac{MCDS}, we do not deactivate these layers, but instead execute the network $k$ times for a given input.
The $k$ outputs generate a distribution $D_{MC}$ of possible solutions $(\theta_i,\phi_i)_{i=1,\ldots,k}$. We therefore stochastically define our solution as $(\mu_{\theta_i},\mu_{\phi_i}) \pm (\sigma_{\theta_i}, \sigma_{\phi_i})$, where
\begin{align}
    (\mu_{\theta_i},\mu_{\phi_i}) ={}& \dfrac{1}{k}\sum_{n=1}^{k} \CNN_{n}(\vec{V}_{i}, L_{i})
    \label{eq:mean} \\
\intertext{represents the predictive posterior mean of our measure, and}
     (\sigma_{\theta_i}, \sigma_{\phi_i}) ={}&  \dfrac{1}{k}\sum_{n=1}^{k} \Big[\CNN_{n}(\vec{V}_{i}, L_{i}) - (\mu_{\theta_i},\mu_{\phi_i})\Big]^2,
     \label{eq:uncertainty}
\end{align}
the uncertainty of our measure. $\CNN_n$ is the \ac{CNN} with active dropout at iteration $n$, $\vec{V}_i$, is the input patch, and $L_i$, is the pivot landmark point $P_i$ normalized in $[0, 1]^3$. It is important to note that, for our implementation, \ac{MCDS} only makes use of dropout layers, and therefore the encoder uncertainty is only partially estimated when considering the uncertainty of the fully connected layers.
The reason for this definition will become clearer in the results section. 
\paragraph*{\Ac{INS}} Instead of observing the variability generated by the dropout layers, at iteration $n$, we iteratively sample a seed point $P_i^n$ from the plane ${\Pi}_i^{n-1}$ predicted at iteration $n-1$ (\autoref{fig:iterative_sampling}), $P_i^1$ being the initial pivot point. 
We sample this point from a uniform distribution, as done for data augmentation, because the network is supposed to be robust to this minimal translation.
From each seed point $P_i^n$, we extract a patch volume $\vec{V}_i^n$, which is centered at this point. 
We assume that an ideal predictor will generate the same coordinates $(\theta, \phi)$ for any of these points.
Again, $n$ iterations will generate a distribution $D_{IN}$ of $n$ possible solutions $(\theta_i,\phi_i)_{i=1,\ldots,n}$, if we redefine the predictive posterior mean and the uncertainty $(\mu_{\theta_i},\mu_{\phi_i}) \pm (\sigma_{\theta_i}, \sigma_{\phi_i})$ as 
\begin{align}
    (\mu_{\theta_i},\mu_{\phi_i}) ={}& \dfrac{1}{k}\sum_{n=1}^{k} \CNN(\vec{V}_i^n, L_{i}^n),
    \label{eq:mean_ins} \\
    \intertext{and}
     (\sigma_{\theta_i}, \sigma_{\phi_i}) ={}& \dfrac{1}{k}\sum_{n=1}^{k} \Big[\CNN(\vec{V}_{i}^n, L_{i}^n) - (\mu_{\theta_i},\mu_{\phi_i})\Big]^2.
     \label{eq:uncertainty_ins}
\end{align}
Here the CNN model $\CNN$ is static and no dropout deactivations are performed. We use this strategy to quantify and reduce the effect of data uncertainty. 
\paragraph*{\Ac{MCDbS}} Finally, we consider the joint evaluation of encoder and decoder uncertainty. Instead of relying only on the variability introduced by the dropout layers, we leverage the uncertainty provided by the DropBlock layers. We stochastically define our solution $(\sigma_{\theta_i}, \sigma_{\phi_i})$ as in \autoref{eq:mean} and \autoref{eq:uncertainty}, but here both the encoder and the fully connected layers combined are a random functional $\CNN_n$ over the execution step $n$. I.e., both the DropBlock and the Dropout layers are kept active. 

\section{Results}
\label{sec:results}
For the evaluation of our model, we utilize only routinely collected data. All data samples were collected from one single hospital, but randomly processed by $11$ different radiology technicians. In this section, we first describe the available data, the pre-processing steps, and evaluate the agreement of the cross-sectional planes {extracted by different operators}.
Afterwards, we report on the accuracy of our model and the improvements introduced by the uncertainty quantification step.

\begin{figure}[t]
    \centering
    \includegraphics[width=0.88\columnwidth]{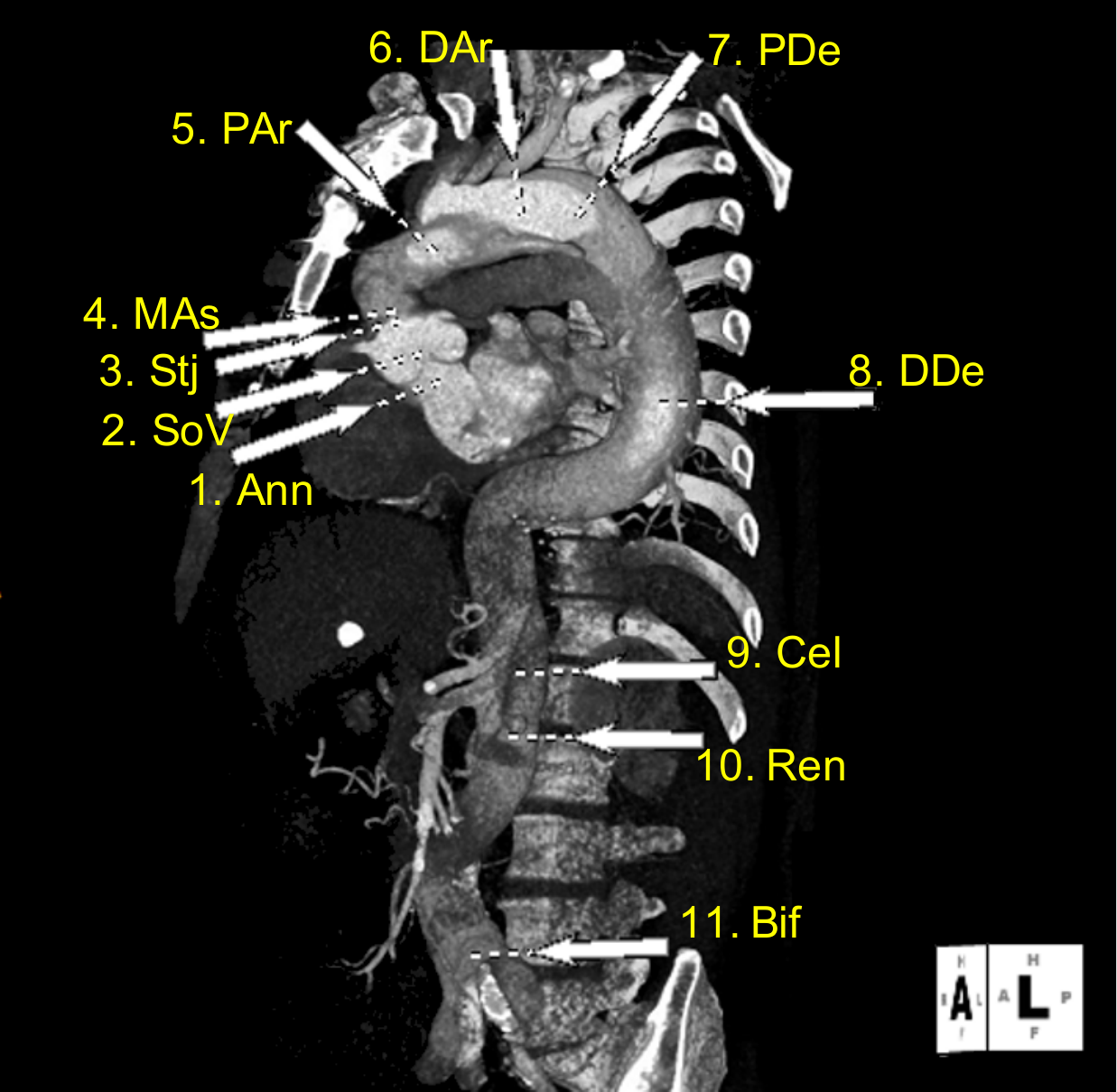}
    \caption{A detailed view of the aortic landmarks at which the measurements were taken. The rendered image shows a case of aortic dissection. The false lumen is clearly distinguishable at the aortic arch. Legend: 1) Aortic annulus (Ann); 2) sinuses of Valsalva (SoV); 3) sinotubular junction (Stj); 4) mid ascending aorta (MAs); 5) proximal aortic arch (PAr); 6) Distal aortic arch (DAr); 7) proximal descending aorta (PDe); 8) distal descending aorta (DDe); 9) aorta at celiac artery (Cel); 10) aorta at inferior main renal artery (Ren); 11) just above iliac bifurcation (Bif). Color image available online.}
    \label{fig:landmarks}
\end{figure}

\subsection{Data acquisition and reference standard}
In total, $162$ {\ac{CTA}} volumes were acquired from $147$ anonymized {\ac{AD}} patients undergoing imaging investigation for aortic surveillance purposes. The volumes were of size $512\times512\times Z$ voxels, with $Z = 910 \pm 170$ slices, and an average spacing of $[0.7, 0.7, 0.7]$\,mm. Each volume was routinely processed by one out of 11 trained operators between May and December 2020. 
For each \ac{CTA} volume, the random operator extracted the relevant cross sections at predefined anatomic locations along the aorta, resulting in $3273$ distinct measurements. The anatomical location of predefined aortic landmarks is shown in \autoref{fig:landmarks}, together with a description of their acronyms. As we deal with clinical annotations, some volumes may present additional measurements or lack some of the defined measurements.
Each cross-sectional plane was extracted manually using a double-oblique reformation tool. All manual processing was performed using state-of-the-art commercial medical image processing software (Aquarius iNtuition, TeraRecon, Inc).
To provide a quantification of the interoperator variability, 12~\ac{CTA} volumes were processed three times each, by three different operators. We asked the operators to specifically measure the aortic diameters at all the locations of interest (\autoref{fig:landmarks}). On average, an experienced operator needed $11$\,s to locate the landmark and $35\textup{--}60$\,s to perform the double-oblique reformation, for each landmark. As a comparison, for healthy aortae, $10$\,s were sufficient to perform the manual reformation. Furthermore, we also i) segmented these 12 CTA volumes using the trained \ac{CNN} provided by \cite{hahn_2020} and ii) additionally tracked the centerlines according to \cite{Krissian_2014}, which allowed us to retrieve the overall aortic centerline for two further methodological references.

\subsection{Pre-processing}
The annotations were initially saved with the proprietary format of Aquarius iNtuition. For each measurement location $i$, we extracted the measurement plane $\Pi_i$ previously selected by the operator using double-oblique reformation.
Each plane $\Pi_i$ is individually defined by a landmark point $P_i \in \mathbb{R}^3$ and its normal vector $\vec{n}_i \in \mathbb{R}^3$.
Landmark point $P_i$ was chosen by the operator as the pivot of the measurement plane.
For process automation, the estimation of the unit normal vector $\vec{n}_i$ would generally require a regression of the three variables $(x, y, z)$.
To reduce complexity, we use spherical coordinates $({r=1}, \theta, \phi)$.
This representation simplifies the problem to the estimation of the two variables {$(\theta, \phi) \in [0,\pi]\times[-\frac{\pi}{2},\frac{\pi}{2}]$}. 
For training and testing of our model, we {shift $\theta$ to the range $[-\frac{\pi}{2},\frac{\pi}{2}]$ and then} normalize the voxel intensity values, as well as $\theta$ and $\phi$, to the range $[-1, 1]$.

%

\pgfplotsset{compat=1.12}
\pgfplotstableread[col sep=comma]{dataPos.dat}\mydataPos
\begin{figure*}

\begin{tikzpicture}
\begin{axis}[width=\linewidth, height=6cm,
  legend style={legend columns=-1},
  legend to name={thelegend},
  name={theaxis},
  ylabel=Limits of agreement,
  label style={font=\small},
  xtick=data,
  xticklabels from table={\mydataPos}{subject},
  xticklabel style={font=\small}, 
  nodes near coords,
  nodes near coords style={font=\tiny, color=black},
  ybar,
  ymin=0,ymax=40,
  ytick={0,5,10,15,20,25,30,35,40},
  yticklabel style={
        /pgf/number format/fixed,
        /pgf/number format/precision=1
}]
\addplot [color=gray!30,fill, bar width=0.0122\linewidth] table [x expr=\coordindex, y={Landmark}] \mydataPos;
\addplot [color=gray!60,fill, bar width=0.0122\linewidth] table [x expr=\coordindex, y={Phi}] \mydataPos;
\addplot [color=gray!90,fill, bar width=0.0122\linewidth] table [x expr=\coordindex, y={Theta}] \mydataPos;
\legend{Landmark (mm) , Phi (deg) , Theta (deg)}
\end{axis}
\node [below] at (theaxis.below south) {\ref{thelegend}};
\end{tikzpicture}
\caption{Limits of agreement of the interoperator variability for the placement of the clinical landmarks and the orientations of the cross-sectional views. All limits were calculated according to \cite{Jones_2011}. We do not show the bias as this is intrisically zero. \emph{All} refers to the overall variability among all landmarks and operators. *) Higher variabilities for PAr, DAr, PDe and, particularly, DDe were expected due to their rather ill-defined nature. For a description of the acronyms, please refer to \autoref{fig:landmarks}. }
\label{fig:landmark_pos}
\end{figure*}
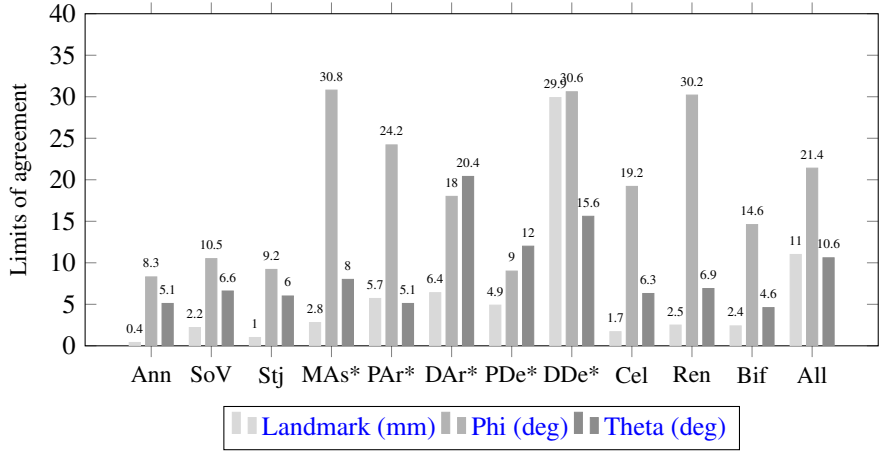

\begin{table}[tb]
    \centering
    \caption{%
       Comparison of the regularization methods. A) Baseline: Dropout regularization is applied only to the fully connected layers; b) Enc. Dropout: Dropout regularization is applied also to the convolutional layers; c) DropBlock: DropBlock regularization is applied to the convolutional layers and Dropout regularization to the fully connected layers. MAE: Mean absolute error.
    }
    \label{tab:ablation}
    \tabulinesep=_3pt^3pt
    \begin{tabu} to \columnwidth {X[4cp]X[2cp]X[2cp]X[4cp]X[4cp]}
        \tabucline[0.10em black]-
        \textbf{CNN Config} &  {Huber loss Train~set} & {Huber loss Valid.~set} & $\theta$-MAE Valid. set [deg] & $\phi$-MAE Valid. set [deg]\\
        \tabucline[0.10em black]-
        \textbf{a)~Baseline} & ${0.003}$ & $0.019$ & $5.75\pm1.03$ & ${15.74\pm1.93}$  \\
         \tabucline[0.05em black!30]-
        \textbf{b)~(a) + Enc. Dropout } & $0.002$ & $0.020$ & $6.65\pm0.98$ & $16.01\pm1.90$ \\  
        \tabucline[0.05em black!30]-
        \textbf{c)~(a)~+ DropBlock} & $0.004$ & ${0.019}$ & $6.35\pm\mathbf{0.70}$  & $15.72\pm\mathbf{1.20}$  \\
        \tabucline[0.10em black]-
    \end{tabu}
\end{table}

\begin{table*}
    \centering
    \caption{%
       Evaluation of the plain CNN predictions on the 12 test volumes with segmentation and interoperator variability. All errors are averaged among the three operators. For a comparison with the available literature, we also compare the centerline approaches of \cite{hahn_2020} (Hahn) and \cite{Krissian_2014} (Krissian) against the three operators. The clinical landmarks marked with * are susceptible to a higher variability due to their ill-posed nature, as shown also in \autoref{fig:landmark_pos}. 
    }
    \begin{minipage}[t]{0.95\columnwidth}
    \tabulinesep=_2pt^2pt
    \label{tab:cnn_centerline_comp}

    \begin{tabu} to \columnwidth
    {X[2cp]X[1cp]X[3cp]X[3cp]}
        \tabucline[0.10em black]-
        \textbf{Landmark} & \textbf{Method} & $\theta$-MAE [deg] & $\phi$-MAE [deg] \\
        \tabucline[0.10em black]-
        Aortic & \textbf{NoUQ}  & $\mathbf{4.38}$ & $\mathbf{6.94}$ \\
        Annulus        & Krissian & $21.80$ & $26.08$ \\       
        (Ann)       & Hahn & N/A & N/A\\
        \tabucline[0.05em black!30]-
         Sinuses of & \textbf{NoUQ} & $\mathbf{3.32}$ & $\mathbf{9.07}$ \\
         Valsalva       & Krissian & $18.09$ & $34.80$\\ 
        (SoV) & Hahn & $11.51$ & $19.68$\\
        \tabucline[0.05em black!30]-
        Sinotubular & \textbf{NoUQ} & $\mathbf{2.30}$ & $\mathbf{11.74}$ \\
          Junction      & Krissian & $19.41$ & $21.24$\\ 
         (Stj)    & Hahn & $9.05$ & $19.46$\\
        \tabucline[0.05em black!30]-
        Mid  & \textbf{NoUQ} & $\mathbf{4.87}$ & $\mathbf{15.31}$ \\
          Ascending      & Krissian & $14.96$ & $19.05$\\ 
        A. (MAs)    & Hahn & $12.64$ & $21.72$\\
        \tabucline[0.05em black!30]-
        Proximal   & \textbf{NoUQ} &  $\mathbf{6.24}$ & $17.86$ \\
         Arch          & Krissian & $18.29$ & $18.59$\\ 
        (PAr)    & Hahn & $10.38$ & $\mathbf{17.34}$\\
        \tabucline[0.05em black!30]-
        Distal & NoUQ & $12.55$ & $10.23$ \\
        Arch & \textbf{Krissian} & $10.74$ & $\mathbf{7.45}$\\ 
        (DAr) & Hahn & $\mathbf{9.36}$ & $10.53$\\
\tabucline[0.05em black!30]-
        Proximal & NoUQ & $11.67$ & $21.81$ \\
        Descending       & Krissian & $10.86$ & $7.74$\\ 
        A. (PDe)* & \textbf{Hahn} & $\mathbf{9.21}$ & $\mathbf{6.30}$\\
        \tabucline[0.05em black!30]-
        Distal & NoUQ  & $6.62$ & $19.82$ \\
        Descending        & Krissian & $9.57$ & $16.22$\\ 
        A. (DDe)* & \textbf{Hahn} & $\mathbf{5.68}$ & $\mathbf{9.99}$\\
        \tabucline[0.05em black!30]-
        A. at celiac & NoUQ & $\mathbf{3.89}$ & $33.95$ \\
        artery & Krissian & $13.52$ & $34.37$\\ 
        (Cel) & \textbf{Hahn} & $4.99$ & $\mathbf{16.44}$\\
        \tabucline[0.05em black!30]-
        A. at inf. &  \textbf{NoUQ}  & $\mathbf{3.93}$ & $26.86$ \\
        main renal & Krissian & $7.07$ & $\mathbf{22.21}$\\ 
        art. (Ren) & {Hahn}  & $4.82$ & ${24.30}$\\
        \tabucline[0.05em black!30]-
        Above &  \textbf{NoUQ} & $\mathbf{4.03}$ & $\mathbf{9.35}$ \\
        iliac bifur. & Krissian* & $11.21$ & $23.05$\\ 
        (Bif) & Hahn & $11.19$ & $20.61$\\
        \tabucline[0.05em black!30]-
        \textbf{Overall}    & \textbf{NoUQ} &  $\mathbf{5.80}$ & $\mathbf{16.63}$ \\
        (All) & Krissian* & $14.21$ & $21.01$\\ 
         & Hahn  & $8.88$ & $16.64$\\
        \tabucline[0.10em black]-

    \end{tabu}
    \end{minipage}
\end{table*}
\subsection{Data analysis} 
\label{sec:data_analysis}
We analyzed the interoperator variability of \emph{three} operators on the {same} subset of 12~\ac{CTA} volumes, which were also segmented. For this purpose, we used the Jones method~(\cite{Jones_2011}), which defines the variance of $m$ measurements performed by $n$ operators as
\begin{align}
    \sigma^2 ={}& \frac{1}{n-1}\sum_{i=1}^{n}\Bigg[\dfrac{1}{m-1} \sum_{j=1}^{m}\big(d_{ij}-\overline{d_i}\big)^2\Bigg],
    \label{eq:variability}
\end{align}
where $d_{ij}$ is the difference between the $j$-th measurement of operator $i$ and the mean difference for $j$, and $\overline{d_i}$ is the mean difference for operator $i$.
The $95\%$ \ac{LOA} with the mean are estimated as LOA~$= \pm1.96 \sigma$. These limits show the uncertainty range which includes $95\%$ of the measurements. We set $n=3$ as the number of operators. However, it can be shown that the \ac{LOA} are equivalent to the more common Bland-Altman limits for the case of $n=2$ operators~(\cite{Jones_2011}).

We computed the \ac{LOA} for the landmark positioning, the angles $(\theta,\phi)$, and the major-axis diameters. The overall \ac{LOA} for the landmark positioning were $\pm10.96$\,mm (\autoref{fig:landmark_pos}). In a detailed analysis, we saw that anatomically well-defined landmarks, such as \textit{Ann}, \textit{Stj} and \textit{Cel}, showed \ac{LOA} between $\pm0.44$\,mm and $\pm2.80$\,mm, while rather ill-defined landmarks, such as \textit{DAr}, \textit{PDe},  and \textit{DDe}, showed \ac{LOA} of up to $\pm6.44$\,mm, $\pm4.88$\,mm, and $\pm29.94$\,mm, respectively. The actual position of these landmarks is shown in \autoref{fig:landmarks}. Although clinically relevant, these landmarks are only qualitatively defined as proximal or distal to a reference anatomical structure, which motivates the larger \ac{LOA}. We chose not to remove these landmarks from our evaluation in order to see the change in performance in relation to the user's accuracy. The major-axis diameters showed \ac{LOA} of $\pm5.33$\,mm, in line with ranges reported in the literature~(\cite{Nienaber_2016}). The \ac{LOA} decreased to $\pm2.90$\,mm when DDe was excluded, which shows the general robustness of the manual annotation. 
For the two angles $(\theta,\phi)$, the \ac{LOA} are reported in \autoref{fig:landmark_pos} for each landmark and combined. In particular, it can be seen how the determination of $\phi$ constitutes a less reproducible task than the determination of $\theta$. The overall \ac{LOA} were $\pm21.39$° and $\pm10.60$°, respectively. We show the impact of this variability on our model in the next sections.

\subsection{Regression study}

After data collection and analysis, we trained our model on $127$ \ac{CTA} volumes and a total of $2556$ measurements. We used the remaining $35$ volumes for validation (27) and testing (12). The testing was conducted on the same $12$ volumes which were annotated by three experts. Each measurement sample was augmented during training by translating the pivot point (landmark) as previously described.
The application was implemented using PyTorch v1.5 and Python v3.7.6. The training was performed on a Linux workstation (GPU: $2\times$ NVidia GeForce RTX 2080Ti, CPU: AMD Threadripper 3960X, 128GB RAM).
The training was terminated using an early-stopping approach after 47 epochs for the baseline model, 61 epochs for the dropout approach and 142 epochs for the DropBlock approach. As expected, the regularization affects the duration of the training. 
\autoref{tab:ablation} shows the effect of two different regularization techniques compared to the baseline \ac{CNN}, where the dropout layers are only placed near the fully connected layers. The regularizations, and particularly DropBlock, appear to reduce the dynamic error related to the predictions. This error is of particular interest in this study, given the focus on reproducibility. For a more comprehensive overview, \autoref{tab:cnn_centerline_comp} shows the prediction errors on a per-landmark basis as well as the overall error on the test set. 
All errors are averaged among the ground truths from the three operators. For comparison, we also report the errors obtained with the methods from \cite{hahn_2020} and \cite{Krissian_2014}. We also need to state that: i) \cite{hahn_2020} can require a manual refinement of the segmentation and ii) \cite{Krissian_2014} requires tuning of different hyperparameters, both leading to an increase in operation times. 

\begin{table}[tb]
    \centering
    \caption{%
       Evaluation of the different execution strategies for uncertainty quantification. Each execution strategy (ES) is evaluated over a different number of iterations $k$.
       As a comparison, NoUQ reports the mean absolute errors for $k=1$, where no uncertainty quantification is performed. 
       The values $\mu\pm\sigma$ correspond to the mean of MAE and standard deviation per measure calculated using the specific execution strategy. Time shows the average execution time for one test sample. For \ac{MCDS}, only the fully connected layers are re-executed, providing lower execution time than \ac{MCDbS}.
    }
    \label{tab:iterative_network_res}
    \tabulinesep=_3pt^3pt
    \begin{tabu} to \columnwidth {X[0.45cp]X[1.5cp]X[4.8cp]X[3.9cp]X[4cp]X[0.8cp]}
        \tabucline[0.10em black]-
        $\mathbf{k}$ & \textbf{ES} & Huber loss & $\theta$-MAE [deg] & $\phi$-MAE [deg] & Time [s]\\
        \tabucline[0.10em black]-
        \textbf{1} & NoUQ & $0.015$ & $7.48$ & $12.40$ &  $0.004$\\
        \tabucline[0.05em black!30]-
        \textbf{5} & \ac{INS} & $0.018\pm0.003$ & $6.35 \pm 1.10$ & $15.52 \pm 1.80$ & $0.24$ \\
          & \ac{MCDS} & $0.016 \pm 0.003$ & $6.58 \pm 1.30$ & $13.97 \pm 2.35$ & $0.01$\\
          & \ac{MCDbS} & $0.014 \pm 0.004$ & $5.44 \pm 1.42$ & $12.02 \pm 2.98$ & $0.02$\\
        \tabucline[0.05em black!30]-
        \textbf{10} & \ac{INS} & $0.017\pm0.003$ & $6.55 \pm 0.89$ & $14.75 \pm 1.81$ & $0.48$ \\
           & \ac{MCDS} & $0.016 \pm 0.003$ & $6.56 \pm 1.43$ & $14.56 \pm 2.57$ & $0.01$\\
           & \ac{MCDbS} & $0.015 \pm 0.005$ & $5.43 \pm 1.64$ & $12.29 \pm 3.22$ & $0.04$\\
        \tabucline[0.05em black!30]-
        \textbf{40} & \ac{INS} & $0.017 \pm 0.003$ & $6.54 \pm 0.92$ & $14.84 \pm 1.78$ & $1.90$ \\
           & \ac{MCDS} & $0.016 \pm 0.004$ & $6.57 \pm 1.52$ & $14.03 \pm 2.89$ & $0.04$\\
           & \ac{MCDbS} & $0.015 \pm 0.005$ & $5.45 \pm 1.74$ & $12.28 \pm 3.58$ & $0.12$\\
        \tabucline[0.05em black!30]-
        \textbf{90} & \ac{INS} & $0.015\pm0.001$ & $7.49 \pm 0.32$ & $12.62 \pm 0.67$ & $4.09$ \\
             & \ac{MCDS} & $0.016 \pm 0.004$ & $6.58 \pm 1.52$ & $14.09 \pm 2.87$ & $0.08$ \\
           & \ac{MCDbS} & $0.014 \pm 0.005$ & $5.42 \pm 1.76$ & $12.12 \pm 3.66$ & $0.26$\\
        \tabucline[0.10em black]-

    \end{tabu}
\end{table}

In both cases, the process took between $10$ and $25$ minutes per volume, excluding landmark selection. 
Additionally, to compare the reproducibility of the plane selection obtained with three approaches (manual, centerline, and \ac{CNN}), we measured the $95\%$ limits of agreement with the mean~(\cite{Jones_2011}). The limits are shown in \autoref{fig:angle_loa} for each landmark and overall. For the centerline~(\cite{hahn_2020}) and CNN (ours) approaches, we used the same landmark positions, which were previously chosen for the manual double-oblique approach -- this to avoid any bias due to temporal task repetition or a different user interface than the clinically approved software. It can be seen how both the centerline method and the \ac{CNN} method provide a higher reproducibility of the plane selection, with the \ac{CNN} approach providing lower overall limits of agreement of $\approx-27\%$ per angle compared to \cite{hahn_2020} and $\approx-52\%$ compared to the manual annotations. 

\subsection{Evaluation of uncertainty}
In \autoref{tab:iterative_network_res}, we report a comparison of the three execution approaches for different numbers of executions $k$ over all the $35$ volumes. In particular, the table shows the average time needed to compute the uncertainty for one sample. The first line, $k=1$, reports the average error and execution time without uncertainty quantification (NoUQ). \autoref{tab:cnn_centerline_res} shows a detailed comparison of the execution approaches for each landmark. The errors are averaged over the annotations of the three operators. \autoref{fig:mc_k_samples} shows the overall effect of iterative sampling for all landmarks from the point of view of reproducibility. In comparison to \autoref{fig:angle_loa}, iterative sampling provides a considerable improvement on $\phi$, which is characterized by a higher interoperator variability in the training data, as previously shown.


\pgfplotsset{compat=1.12}
\pgfplotstableread[col sep=comma]{data.dat}\mydata
\pgfplotstableread[col sep=comma]{dataK.dat}\mydataK

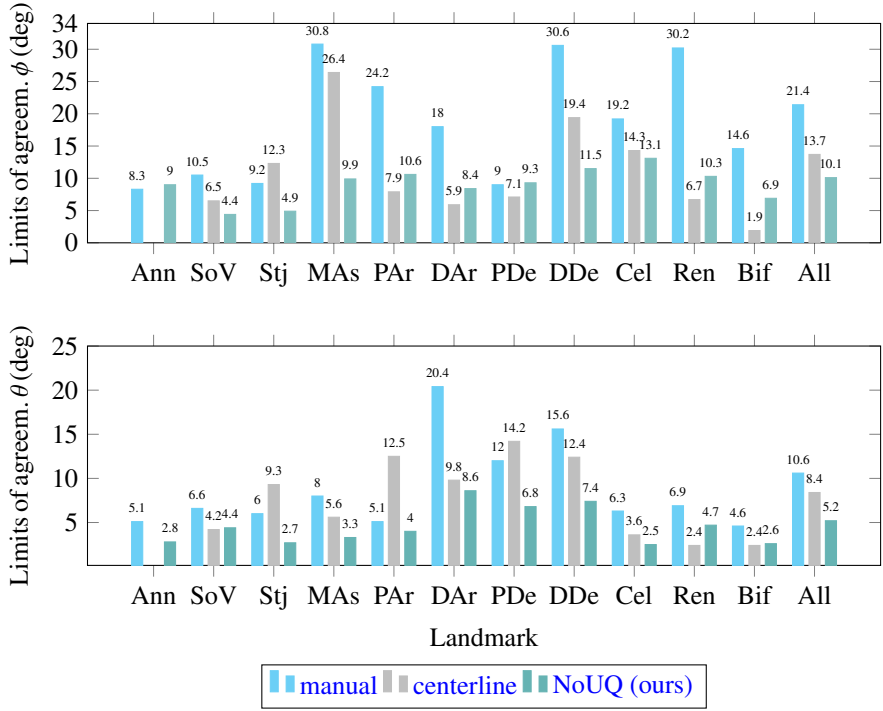
\begin{figure*}
\begin{tikzpicture}
\begin{axis}[width=\linewidth, height=4.5cm,
  legend style={legend columns=-1},
  legend to name={thelegend},
  name={theaxis},
  ylabel={Limits\,of\,agreem.\,{$\mathbf{\phi}$}\,(deg)},
  xtick=data,
  xticklabels from table={\mydata}{subject},
  ybar,
  ymin=0,ymax=34,
  nodes near coords,
  nodes near coords style={font=\tiny, color=black},
  ytick={0,5,10,15,20,25,30,34},
  yticklabel style={
        /pgf/number format/fixed,
        /pgf/number format/precision=2
}]
\addplot [color=cyan!50,fill, bar width=0.0122\linewidth] table [x expr=\coordindex, y={Phi}] \mydata;
\addplot [color=lightgray,fill, bar width=0.0122\linewidth] table [x expr=\coordindex, y={PhiCL}] \mydata;
\addplot [color=teal!50,fill, bar width=0.0122\linewidth] table [x expr=\coordindex, y={PhiCNN}] \mydata;
\end{axis}
\node [below] at (theaxis.below south) {\ref{thelegend}};
\end{tikzpicture}
\begin{tikzpicture}
\begin{axis}[width=\linewidth, height=4.5cm,
  legend style={legend columns=-1},
  legend to name={thelegend},
  name={theaxis},
  xlabel=Landmark,
  ylabel=Limits\,of\,agreem.\,$\mathbf{\theta}$\,(deg),
  xtick=data,
  xticklabels from table={\mydata}{subject},
  ybar,
  ymin=,ymax=25,
  nodes near coords,
  nodes near coords style={font=\tiny, color=black},
  ytick={0,5,10,15,20,25},
  yticklabel style={
        /pgf/number format/fixed,
        /pgf/number format/precision=2
}]
\addplot [color=cyan!50,fill, bar width=0.0122\linewidth] table [x expr=\coordindex, y={Theta}] \mydata;
\addplot [color=lightgray,fill, bar width=0.0122\linewidth] table [x expr=\coordindex, y={ThetaCL}] \mydata;
\addplot [color=teal!60,fill, bar width=0.0122\linewidth] table [x expr=\coordindex, y={ThetaCNN}] \mydata;
\legend{manual, centerline, NoUQ (ours)}
\end{axis}
\node [below] at (theaxis.below south) {\ref{thelegend}};
\end{tikzpicture}

\caption{Limits of agreement (95\%) of the interoperator variability for the two angles $(\theta,\phi)$ according to \cite{Jones_2011} (absolute values, \textbf{lower values are better}). For better readability, we do not report biases: these are zero in all cases, given we measure the variability against the mean. \emph{All} refers to the overall variability among landmarks and operators. Lower limits of agreement correspond to a higher reproducibility of the results. For each landmark, \emph{manual} refers to double-oblique reformation, \emph{centerline} to \cite{hahn_2020}, and \emph{NoUQ} to our method without uncertainty quantification (UQ). The effect of UQ is reported separately.}
\label{fig:angle_loa}
\end{figure*}


\pgfplotsset{compat=1.12}
\pgfplotstableread[col sep=comma]{data.dat}\mydata
\pgfplotstableread[col sep=comma]{dataK.dat}\mydataK

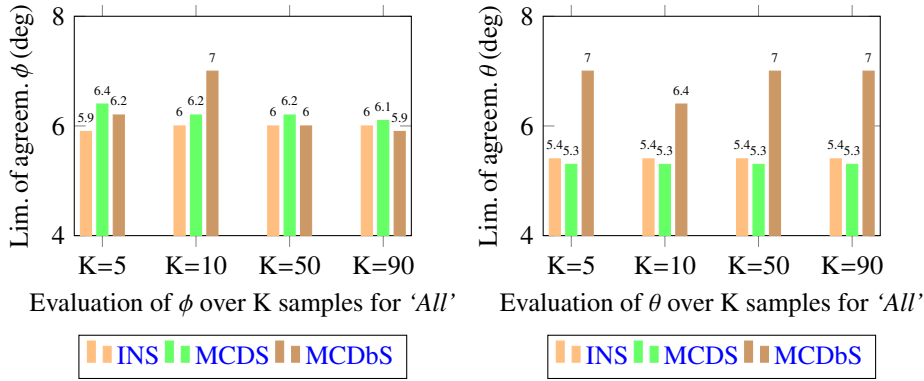
\begin{figure*}

\begin{tikzpicture}
\begin{axis}[width=0.5\linewidth, height=4.5cm,
  legend style={legend columns=-1},
  legend to name={thelegend},
  name={theaxis},
  xlabel={Evaluation of $\phi$ over K samples for \textit{`All'}},
  ylabel=Lim.\,of\,agreem.\,{$\mathbf{\phi}$}\,(deg),
  xtick=data,
  xticklabels from table={\mydataK}{subject},
  ybar,
  ymin=4,ymax=8,
  nodes near coords,
  nodes near coords style={font=\tiny, color=black},
  ytick={4,6,8},
  yticklabel style={
        /pgf/number format/fixed,
        /pgf/number format/precision=2
}]
\addplot [color=orange!50,fill, bar width=0.0122\linewidth] table [x expr=\coordindex, y={INSP}] \mydataK;
\addplot [color=green!60,fill, bar width=0.0122\linewidth] table [x expr=\coordindex, y={MCDSP}] \mydataK;
\addplot [color=brown!80,fill, bar width=0.0122\linewidth] table [x expr=\coordindex, y={MCDBSP}] \mydataK;
\legend{INS, MCDS, MCDbS}

\end{axis}
\node [below] at (theaxis.below south) {\ref{thelegend}};
\end{tikzpicture}
\begin{tikzpicture}
\begin{axis}[width=0.5\linewidth, height=4.5cm,
  legend style={legend columns=-1},
  legend to name={thelegend},
  name={theaxis},
  xlabel=Evaluation of $\theta$ over K samples for \textit{`All'},
  ylabel=Lim.\,of\,agreem.\,$\mathbf{\theta}$\,(deg),
  xtick=data,
  xticklabels from table={\mydataK}{subject},
  ybar,
  ymin=4,ymax=8,
  nodes near coords,
  nodes near coords style={font=\tiny, color=black},
  ytick={4,6,8},
  yticklabel style={
        /pgf/number format/fixed,
        /pgf/number format/precision=2
}]
\addplot [color=orange!50,fill, bar width=0.0122\linewidth] table [x expr=\coordindex, y={INST}] \mydataK;
\addplot [color=green!60,fill, bar width=0.0122\linewidth] table [x expr=\coordindex, y={MCDST}] \mydataK;
\addplot [color=brown!80,fill, bar width=0.0122\linewidth] table [x expr=\coordindex, y={MCDBST}] \mydataK;
\legend{INS, MCDS, MCDbS}
\end{axis}
\node [below] at (theaxis.below south) {\ref{thelegend}};
\end{tikzpicture}

\caption{Limits of agreement (95\%) of the interoperator variability for the two angles $(\theta,\phi)$ according to \cite{Jones_2011} (absolute values, lower values are better). We report the agreement improvement for the overall category after applying the three different uncertainty quantification strategies. Results are to be compared with \autoref{fig:angle_loa}.}
\label{fig:angle_loa_uq}
\end{figure*}
\begin{table*}
    \centering
    \caption{%
       Evaluation of the different approaches on the 12 test volumes with segmentation and interoperator variability. We provide a comparison of the different execution strategies against the three operators. The clinical landmarks marked with * are susceptible to a higher variability due to their ill-posed nature, as shown also in \autoref{fig:landmark_pos}. 
    }
    \begin{minipage}[t]{0.95\columnwidth}
    \tabulinesep=_2pt^2pt
    \label{tab:cnn_centerline_res}

    \begin{tabu} to \columnwidth
    {X[2cp]X[1cp]X[3cp]X[3cp]}
        \tabucline[0.10em black]-
        \textbf{Landmark} & \textbf{ES} & $\theta$-MAE [deg] & $\phi$-MAE [deg] \\
        \tabucline[0.10em black]-
        Aortic & \ac{INS} & $4.38\pm0.17$ & $6.92\pm0.47$ \\
        Annulus      & \ac{MCDS}  & $4.37 \pm 1.37$ & $6.88 \pm 2.34$  \\
        (Ann) & \textbf{\ac{MCDbS}} &  $\mathbf{3.00 \pm 1.24}$ & $\mathbf{4.95 \pm 1.87}$  \\
        \tabucline[0.05em black!30]-
        Sinuses  of & \ac{INS} & $\mathbf{3.30\pm0.15}$ & $9.06\pm0.50$ \\
        Valsalva   & \ac{MCDS} & ${3.31 \pm 1.29}$ & $9.23 \pm 2.42$ \\
        (SoV)  & \textbf{\ac{MCDbS}} & $3.49 \pm 1.28$ & $\mathbf{6.28 \pm 1.61}$ \\
        \tabucline[0.05em black!30]-
        Sinotubular  & \ac{INS} & $\mathbf{2.29\pm0.17}$  & $11.82\pm0.48$  \\
        Junction   & \ac{MCDS} & $2.55 \pm 1.22$ & $11.91 \pm 2.50$ \\
        (Stj)   & \textbf{\ac{MCDbS}} & $3.02 \pm 1.13$ & $\mathbf{4.84 \pm 1.87}$ \\
        \tabucline[0.05em black!30]-
        Mid & \ac{INS} & $\mathbf{4.87\pm0.15}$  & $15.52\pm0.82$  \\
        Ascending   & \ac{MCDS} & $4.90 \pm 1.29$ & $15.40 \pm 3.07$ \\
        A. (MAs)   & \textbf{\ac{MCDbS}} & $5.47 \pm 1.61$ & $\mathbf{9.13 \pm 2.32}$ \\
        \tabucline[0.05em black!30]-
        Proximal & \ac{INS} &  $6.25\pm0.17$ & $17.69\pm0.61$ \\
        Arch   & \ac{MCDS} & $6.52 \pm 1.27$ & $17.92 \pm 2.45$ \\
        (PAr)   &\textbf{\ac{MCDbS}} &  $\mathbf{4.75 \pm 1.31}$ & $\mathbf{9.97 \pm 1.82}$ \\
        \tabucline[0.05em black!30]-
        Distal & \ac{INS} &  $12.54\pm0.49$ & $10.30\pm0.41$ \\
        Arch\textbf{*}   & \ac{MCDS} &  $12.69 \pm 2.04$ & $10.28 \pm 1.89$ \\
        (DAr)    & \textbf{\ac{MCDbS}} &  $\mathbf{9.32 \pm 2.09}$ & $\mathbf{8.64 \pm 1.88}$ \\
\tabucline[0.05em black!30]-
        Proximal & \ac{INS} & $11.66\pm0.52$ & $21.71\pm0.71$ \\
        Descending   & \ac{MCDS} & $11.91 \pm 2.08$ & $21.81 \pm 3.10$ \\
        A. (PDe)*   & \textbf{\ac{MCDbS}} & $\mathbf{9.53 \pm 2.11}$ & $\mathbf{19.44 \pm 2.90}$ \\
        \tabucline[0.05em black!30]-
        Distal & \textbf{\ac{INS}} & $\mathbf{6.62\pm0.30}$ & $20.04\pm1.14$ \\
        Descending   & \ac{MCDS} & $6.82 \pm 1.83$ & $19.64 \pm 2.98$ \\
        A. (DDe)\textbf{*} & \textbf{\ac{MCDbS}} & $10.27 \pm 2.33$ & $\mathbf{16.03 \pm 2.95}$ \\
        \tabucline[0.05em black!30]-
        A. at celiac & \ac{INS} & $\mathbf{3.86\pm0.20}$ & $34.08\pm0.86$ \\
        artery   & \ac{MCDS} & $4.09 \pm 1.63$ & $33.83 \pm 3.98$ \\
        (Cel) & $\mathbf{\ac{MCDbS}}$ & $7.61 \pm 2.17$ & $\mathbf{24.34 \pm 3.77}$ \\
        \tabucline[0.05em black!30]-
        A. at inf. & \ac{INS} & $\mathbf{3.89\pm0.15}$ & $26.86\pm0.88$ \\
        main renal  & \ac{MCDS} & $4.07 \pm 1.50$ & $26.50 \pm 3.52$ \\
        art. (Ren) & \ac{MCDbS} & $7.99 \pm 1.93$ & $25.36 \pm 3.90$ \\ 
        \tabucline[0.05em black!30]-
        Above & \ac{INS} & $\mathbf{4.01 \pm 0.13}$ & $9.27\pm0.60$ \\
        iliac bifur. & \ac{MCDS} & $4.47\pm1.15$ & $9.58\pm2.81$\\
        (Bif)  & \textbf{\ac{MCDbS}}  & $4.65 \pm 1.76$ & $\mathbf{7.40 \pm 2.67}$ \\
        \tabucline[0.05em black!30]-
        \textbf{Overall} & \ac{INS} & $\mathbf{5.81 \pm 0.21}$ & $16.62 \pm 0.63$ \\
        (All) & \ac{MCDS} & $5.99 \pm 1.45$ & $16.90\pm2.88$ \\
        & \textbf{\ac{MCDbS}} & $\mathbf{5.31 \pm 2.77}$ & $\mathbf{12.53 \pm 2.65}$ \\
        \tabucline[0.10em black]-
    \end{tabu}
    \end{minipage}
\end{table*}

\section{Discussion}
\label{sec:discuss}

We introduced how surveillance of \ac{AD} patients requires cross-sectional aortic measurements in \ac{CTA} imaging. Although centerline analysis is recommended, this remains a challenging and expensive task for \ac{AD}. 
Hence, manual double-oblique reformation is usually preferred for patients with challenging aortic conditions, like \ac{AD}. This generates a growing amount of valid data in the clinical routine, although the interoperator variability between highly trained operators cannot be ignored. Our \ac{CNN}, trained only on routinely collected data, was able to generalize well even if training and test samples were generated by $11$ different operators. Nonetheless, the accuracy of the model alone does not guarantee model validity~(\cite{Li2020AccuracyReprod}). Therefore, we further examined the robustness of the model against three independent operators and two centerline methods. 
Although centerline methods are a common standard~(\cite{Gamechi_2019}), their extraction in \ac{AD} cases can be expensive or highly inaccurate: \ac{CNN} segmentation methods are extremely expensive to train~(\cite{hahn_2020}) and vessel filters typically generate independent centerlines for the two lumina~(\cite{Krissian_2014}), which is useful for surgery planning~(\cite{zhao2021automatic}) but not for surveillance imaging. We posed this as a regression problem, where we predict the direction of the cross-sectional plane. 
Unlike most machine learning application studies, we did not limit this to an accuracy evaluation, but we further considered the role of uncertainty and validated the model against three independent operators. The remainder of this section provides a more detailed discussion on this matter. 

\subsection{Effect of regularization and execution strategies}
Uncertainty plays a key role when dealing with real-word data. As introduced, a possible approach for model uncertainty quantification is to perform a Monte Carlo sampling of the \ac{CNN} predictions by leaving the dropout layers active also at test time. However, the introduction of DropBlock as regularization technique for the convolutional layers brought a noticeable reduction of uncertainty, especially for the $\phi$ angle, characterized by a higher interoperator variability. 
The evaluation also suggests that the application of dropout solely to the fully connected layers can lead to better performance than applying dropout also to the convolutional layers. 
Although the resulting MAE is not negligible, the low standard deviation of the network shows a high precision, a relevant factor for task reproducibility.
{Finally, we observed an interesting relation between the evaluated approaches and the interoperator variability of the training set: for variables with a relatively lower variability, like $\theta$, \ac{INS} provides often more reliable solutions, characterized by a deviation $\sigma\leq0.20$ with $10$ samples. Variables with higher variability are characterized by an \ac{INS} deviation above this threshold. In this case, \ac{MCDbS} provides a considerably lower error, unless $\sigma\geq2.50$, where the previous centerline method is still more reliable. These thresholds show where the \ac{CNN} lacks sufficient training examples to correctly generalize on test data. \textit{Thanks to the short computation time, both methods can be executed in parallel, and a simple binary threshold can be used to recommend a solution to the user}.} 

\subsection{\ac{CNN} performance and time reduction}
In \autoref{tab:iterative_network_res}, we observe that the average execution time of the neural network is only $4$\,ms for single executions without uncertainty quantification (NoUQ). This is in contrast with timings of up to $60$\,s for the manual reformation and  $30$\,m for the evaluated centerline methods. The relatively short execution time of the \ac{CNN} makes the approach easily applicable in soft real-time applications, such as the interactive delineation of cross-sectional planes. As expected, the \ac{INS} approach shows higher computation times, because the updated volume patch $\mathbf{V}_i^n$ has to be loaded on the GPU at each execution step $n$. Nonetheless, $k=10$ steps appear to be sufficient to quantify the data uncertainty for an overall execution time of $0.48$\,s. The performance increment with a higher $k$ becomes less relevant compared to execution time. For the specific task, data uncertainty appears to play a less important role than model uncertainty (\ac{MCDS} and \ac{MCDbS}). This is in line with the statements from \cite{kendall_2017}, for whom the type of relevant uncertainty depends on the amount of training volumes. Large collections are difficult to find for rare conditions such as \ac{AD}. \ac{MCDS}, where only the uncertainty of the fully connected layer is exploited, shows a low computational cost. This is due to the fact that the output of the encoding layer can be static after the first execution, and only the fully connected layers need to be iteratively re-executed. 
Compared to \ac{INS}, \ac{MCDS} slightly reduces the bias at the cost of higher uncertainty. Finally, \ac{MCDbS} introduces an interesting outcome. For $\theta$, where the interoperator variability is lower (\autoref{fig:angle_loa}), the biases are lower than the other two methods, at the cost of higher uncertainty. For the angle $\phi$, which showed a much higher interoperator variability, the bias reduction is even stronger. This behaviour is more evident in \autoref{tab:cnn_centerline_res}. 
In our empirical tests, hybrid approaches between \ac{INS} and Monte Carlo sampling did not show any improvement on top of the discussed methods. 
All the approaches are shown in detail in \autoref{tab:cnn_centerline_res} and compared to a centerline-base method. 

Overall, the \ac{CNN} approach without uncertainty quantification (NoUQ) shows similar performance to that of the centerline method. A lower error is particularly seen for the angle $\theta$. For this angle, the uncertainty quantification (UQ) strategies introduce small benefits, compared to $\phi$. The second angle benefits considerably from the application of \ac{MCDbS}, which lowers the bias without increasing the uncertainty when compared to \ac{MCDS}. Observing the landmarks, \ac{MCDbS} is beneficial in all the cases for $\phi$, although it can be noticed that its effects are less strong on ill-posed landmarks (marked with *). Altogether, this suggests that \ac{MCDbS} is an efficient UQ strategy when the training data is limited and has a high interoperator variability. For larger collections or, in case of $\theta$, when the interoperator variability is low, \ac{INS} or the state-of-the-art \ac{MCDS} approach have higher benefits.

Furthermore, this approach did not only 
reduce the disagreement but also the 
processing time. {For healthy patients, a radiology technologist required up to $30$\,s to extract one cross-section, while, for \ac{AD} patients this process required up to $60$\,s. This shows how the disease complicates the measurement process. As a comparison, the extraction of smooth centerlines required between $11$\,min and $23$\,min per patient, while the suggested approaches require between $4$\,ms and $0.48$\,s to perform the same task with an estimation of the uncertainty.}

\subsection{Task reproducibility}
A strong motivation for this work is reproducibility. In \autoref{fig:angle_loa}, we depict the $95\%$ limits of agreement for both angles at each landmark position. It can be seen how the agreement between the three observers is higher when the operator only depicts the landmark point and the plane is retrieved either from the vessel centerline or with the CNN. A remark must be made for the aortic annulus (\textit{Ann}): For this landmark, our method cannot be compared to the centerline method of \cite{hahn_2020}: The aortic segmentation does not always correctly include the aortic root, and, due to the thinning process, the centerlines do not cover the extremities of the aorta.
We therefore marked the value as not available (N/A) for the centerline method. 
In general, planes retrieved with the \ac{CNN} method show higher agreement than those retrieved from centerlines, without requiring segmentation and skeletonization, but relying only on the task learned from cheaper, clinical annotations. 
Additionally, for excentric diseased cases, such as dissections and aneurysms, the processed centerlines usually do not match the ideal path, and manual smoothing steps are necessary to enhance the quality of the centerline~(\cite{Egger_2012}). These steps add a further degree of uncertainty to the final result, which cannot be easily quantified. Our method, instead, provides more accurate and reproducible estimations of the planes with their related uncertainty.
The uncertainty quantification plays a key role especially for those variables intrinsically characterized by a higher uncertainty, like $\phi$. The uncertainty quantification methods evaluated in this paper considerably reduce this uncertainty between operators. As shown in \autoref{fig:angle_loa} (3-4), with the introduction of these techniques, both the limits of agreement settle at around 5-6 degree, suggesting that this residual disagreement might be due to the uncertainty in landmark placement.

\section{Conclusion}

In this work, we addressed the clinical problem of retrieving cross-sectional measurement planes of the aorta in \ac{CTA} volumes, with a particular emphasis on aortic dissections. For these cases, the extraction of the cross-sectional planes is a challenging, time-consuming and considerably uncertain task, which is still performed manually in the clinical routine.

Our approach is centerline- and segmentation-independent, which is important for diseased vessels, and extracts the cross-section from a single landmark point. In particular, we showed how our \ac{CNN} model can be used to predict the orientation of a cross-sectional plane with a quantification of the model uncertainty. Additionally, we showed how our method performs in comparison to the interoperator variability of three experienced users and against two existing methods. The model appears to generalize well when compared to the average error of an expert user and shows similar, if not better, performance to one of the centerline methods, which instead requires a segmentation of the vessel. We showed how the suggested approach can reduce the $95\%$ limits of agreement, thus guaranteeing higher reproducibility. 
Additionally, we showed how the quantification of the encoder uncertainty can help dealing with training data with intrinsically higher interoperator variability. 

Future work includes a large-scale, multi-center evaluation of our approach within the clinical routine. In addition, we plan to extend our approach to a broader set of vascular diseases, like carotid artery stenosis~(\cite{carotidStenosis2}), and to automate the placement of the seed points with automatic landmark detection approaches~(\cite{Payer_2019}). Uncertainty visualization methods may also leverage the uncertainty information and guide the radiologists to a range of possible solutions.

\section*{Acknowledgments}
We acknowledge that this work was approved by the Institutional Review Board of Stanford University, School of Medicine (\textit{IRB n. 47939}) and received fundings from the TU Graz LEAD Project \emph{Mechanics, Modeling and Simulation of Aortic Dissection}, the Austrian Science Fund (FWF) \textit{KLI 678-B31} \emph{enFaced}, the REACT-EU project KITE (Plattform für KI-Translation Essen), and the Austrian Marshall Plan Foundation Scholarship. Furthermore, we thank all members of the 3D and Quantitative Imaging Laboratory (Stanford University), Lewis D. Hahn (UC San Diego), and Sascha Ranftl (TU Graz).

\bibliographystyle{model2-names.bst}
\biboptions{authoryear}
\bibliography{refs}



\end{document}